\documentclass[11pt]{article}
\usepackage[a4paper,hmarginratio=1:1,vmarginratio=2:3,
totalwidth=16.5cm,totalheight=21.51cm]{geometry}  

\usepackage{epsfig} 
\newcommand{\cO}{{\cal O}}
\newcommand{\cR}{{\cal R}}
\newcommand{\cW}{{\cal W}}
\newcommand{\bZ}{{\bf Z}}

\newcommand{\ra}{\rightarrow}
\newcounter{oldcounter}
\addtocounter{equation}{1}
\begin{document} 
\begin{flushright}  
{DAMTP-2003-83}\\
\end{flushright}  
\vskip 3 cm 
\begin{center} 
{\Large {\bf Regularisation Techniques for the Radiative Corrections}}\\
\bigskip
 {\Large {\bf of Wilson lines and Kaluza-Klein states}} \\
\vspace{1.2cm} 
{{\bf D.M. Ghilencea}}\\
\vspace{1.cm}
{\it  D.A.M.T.P., C.M.S., University of Cambridge, \\
          Wilberforce Road, Cambridge CB3 OWA, United Kingdom.}\\
\end{center} 
\vspace{1.7cm} 
\begin{abstract}
Within an effective field theory framework we compute the most general
structure of the one-loop corrections to the 4D gauge couplings in 
one- and two-dimensional orbifold compactifications with non-vanishing
constant gauge background (Wilson lines). Although such models are
non-renormalisable, we  keep the analysis general by
considering the  one-loop  corrections in three regularisation schemes: 
dimensional regularisation (DR),  Zeta-function regularisation 
(ZR)  and proper-time cut-off  regularisation (PT).  The 
relations among the results obtained in these  schemes are carefully
addressed. With minimal re-definitions of the parameters involved, 
the results  obtained for the radiative corrections
can be applied to most  orbifold compactifications  with one or     
two compact  dimensions. The link with string theory is discussed.
We  mention a possible implication for the gauge couplings 
unification in such models.

\vspace{0.5cm}
\end{abstract}

\newpage\setcounter{page}{1}
$ $
\vspace{2.5cm}
\tableofcontents{}
\newpage

\section{Introduction.}

There currently  exists great interest in the physics of
compact dimensions in the context of  experimental and 
theoretical efforts to understand the physics beyond the Standard 
Model (SM). Model building beyond the Standard Model is in general 
 based on 
additional assumptions such as  higher amount of symmetry 
(supersymmetry, gauge symmetry), additional compact dimensions, string
theory, etc,  which attempt to explain the physics at high energy
scales and  which must ``recover'' in the low energy limit the Standard
Model  physics. One way to ``relate'' these two very different
energy scales and thus provide an insight into physics beyond the SM is to 
study the behaviour of the gauge couplings of the model by considering 
their one-loop radiative corrections.

In this paper we use  an effective field theory (EFT) approach to 
compute radiative corrections to the 4D gauge couplings induced by
orbifold compactifications with Wilson line background.
Such corrections are related to the ``threshold effects'' 
of Kaluza-Klein (KK) states  associated with the compact 
dimensions. In general higher dimensional  models also have a larger
gauge symmetry than  that in supersymmetric versions of SM-like models.
Examples  of  breaking the higher dimensional gauge symmetries are
the Hosotani \cite{Hosotani:1983xw} or  Wilson line
\cite{Candelas:en,Green}  mechanism  which is natural for manifolds
not simply connected. This symmetry breaking mechanism  affects
the 4D  Kaluza-Klein masses and thus the one-loop corrections to
the gauge couplings.
We thus discuss the corrections 
to  the couplings due to Kaluza-Klein modes in the  presence
of such  symmetry breaking mechanism.

Radiative corrections from compact dimensions  were studied 
in the past in effective field theory approaches 
(see for example \cite{Dienes:1998vg,Ghilencea:1999cy,Oliver:2003cy,
Goldberger:2002cz}) or in  string theory (see for example 
\cite{Ibanez:1986xy,Dixon:1990pc,Ferrara:1991uz,
Nilles:1997vk,Mayr:1995rx,Lust:2003ky}). 
However, on the field theory level the effect of Wilson lines on the 4D gauge
couplings is little explored even for the simplest field theory 
orbifolds, due to technical difficulties, and this motivated the
present work. Such analysis is relevant given the importance of Wilson
lines  for phenomenology.
Further,  field theory calculations are usually performed for a 
particular choice of the regularisation scheme 
and the link with other schemes is not always clear. 
Such link is important because  models with compact dimensions are
non-renormalisable and comparing  the results for radiative 
corrections in various regularisations  provides additional 
information on the UV behaviour of the models.  

Previous  studies  of the  link between field  and string theory 
results \cite{Ghilencea:2002ff,Ghilencea:2002ak,Ghilencea:2003kt}  for 
Kaluza-Klein radiative
corrections  suggest that in some cases the string "prefers" on the 
field theory side a  proper-time cutoff regularisation for the UV region. 
However, such regularisation is not gauge invariant in field theory.
In this context our purpose is  to provide for one- and
two-dimensional field theory compactifications, the most  general
structure of the one loop 
corrections to gauge couplings  in the presence of Wilson lines 
background, in dimensional
regularisation  (DR) and  zeta-function regularisation (ZR). Their
link with results in proper-time cutoff regularisation (PT) and 
in (heterotic) string theory is also discussed.  Our results   for the 
radiative corrections are very general and  can be  easily applied to 
specific models.

The analysis starts from the observation that
while the field content which contributes to the one-loop corrections
is strongly model dependent, the general  structure 
of the mass spectrum of Kaluza-Klein modes is determined 
 by  the (eigenvalues of  the Laplacian $\Delta$ in a constant 
gauge background for the) manifold/orbifold of compactification. 
For the  particular but  often considered cases of an  orbicircle or
two-dimensional orbifold $T^2/Z_N$, the integrals over compact
dimensions and sums over associated non-zero Kaluza-Klein modes 
can be performed in a  model-independent way. Once this is done, 
this leaves the much simpler task of determining the exact values of the 
beta functions to a model-by-model analysis.

More explicitly, note that the general structure of one-loop
corrections to the inverse of the tree level ("bare") 
gauge couplings $\alpha_i$, induced by Kaluza-Klein modes,
 may be  written  {\it formally} as
\begin{eqnarray}\label{frm}
\Omega_i^*= \textrm{tr}\frac{\beta(\sigma)}{4\pi} 
\ln \det \Delta (\sigma)
\end{eqnarray}
$\Delta(\sigma)$ is the (spectrum of the) Laplacian on the 
manifold/orbifold
considered. $\beta(\sigma)$ is the one-loop beta function of a
``component'' state of charge $\sigma$ under some symmetries of
compactification (boundary conditions) or a constant gauge background,
and belonging to a particular  multiplet/representation. 
The trace ``tr'' acts over all states/representations of the theory
which have Kaluza-Klein modes associated. 
In  the string context $\Omega_i^*$ can be related  to
the free energy of compactification \cite{Ferrara:1991uz}, (see also 
\cite{LopesCardoso:1994ik}) and torsion \cite{Nash:sf},
\cite{Friedmann:2002ty}.

In general the dependence of the spectrum of the 
 Laplacian $\Delta$ on the charge  
($\sigma$) prevents one from factorising the  $\sigma$ dependence 
(full beta function) in front of the logarithm (\ref{frm}). 
However, we regard  $\sigma$ as a {\it fixed parameter} and 
compute  $\ln\det \Delta(\sigma)$ in general, for one and two
dimensional orbifolds. Effectively this means to replace 
$\Delta$ by its eigenvalues expressed in some
mass units. In an effective field theory the natural mass unit
is that associated with its ultraviolet cutoff $\Lambda$.
With this argument  eq.(\ref{frm}) gives the usual  sum of
logarithms $\sum_{n}\ln\Lambda/M_{n}(\sigma)$  
known in field theory \cite{IZ}, with $M_{n}(\sigma)$ the 
mass of a  Kaluza-Klein state of level~$n$ 
(for two dimensions $n$ is replaced by a set of two integers
$\{n_1,n_2\}$ associated each with one compact dimension). 
One then multiplies this sum by $\beta(\sigma)$ and performs the 
remaining model-dependent sum (``tr'') over $\sigma$.

In the presence of a constant gauge background/twists (Wilson
lines) the  eigenvalues of the Laplacian are changed by an amount 
function of $\sigma$, related to the Wilson lines vev's. The
correction  of the Wilson lines to the gauge couplings
may  be regarded in some cases as an
additional effect  (``perturbation'') to that due to Kaluza-Klein
modes alone, for vanishing Wilson vev's. 
This idea  may in principle be used for much more complex manifolds 
(for example Calabi Yau, $G_2$ manifolds) with  Wilson  lines background, 
to relate their associated one loop corrections to those for vanishing 
background and the corresponding topological quantities 
(torsion)~\cite{Friedmann:2002ty}.

There remains the  question of the regularisation of (\ref{frm}).
This equation only makes sense in the presence of a regularisation
both in the UV and IR. Indeed,  $\det \Delta$ vanishes for massless
modes  and an IR regulator (mass shift)  $\chi$ is in general required to
ensure $\ln\det\Delta$ is well-defined {\it before} proceeding further. 
Thus one should in fact  compute 
$\ln\det (\Delta-\chi^2)$.  This is  ``avoided'' in the sense  that 
one  usually evaluates only the (IR finite) contribution of the 
{\it massive} 
(Fourier) modes alone, denoted  $\ln\det (\Delta')$.
This means that one implicitly takes the limit $\chi\!\ra\! 0$ in 
the massive 
modes' sector. This leaves the IR  regulator be present
and act   only in the sector of  the massless modes  alone.
Further, the correction  $\ln\det(\Delta')$ requires itself a
regularisation, this time in the UV
\cite{Ghilencea:2002ff,Ghilencea:2002ak} since the contribution of
the  KK tower is in general UV divergent and a regulator  denoted 
$\xi$ ($\xi\! \ra\! 0$) is introduced.
The important point is that the limits $\chi\ra 0$ and $\xi\ra
0$ of  the above UV and IR regularisation of $\ln\det(\Delta'-\!\chi^2)$ 
 do not  necessarily commute in the {\it massive}  modes' sector~!
The two regularisations and the
UV and IR regions may not  be ``decoupled''
from each other and  a UV-IR ``mixing'' (UV divergent, IR finite)
is present. See \cite{Ghilencea:2002ak,toappear} for 
an  example with two  compact dimensions   and its 
 string theory interpretation.   Such situation can arise in
non-renormalisable theories  due to summing over  two 
{\it infinite}-level Kaluza-Klein towers, and  is not 
present if  the two sums are truncated to a  finite number of  modes.
We will encounter  this issue in Section  \ref{two_dim_loop}.

In the following we compute the one-loop corrections due to {\it massive}
modes to the 4D gauge couplings for one and two dimensional
orbifolds, in the DR, ZR and PT  regularisation schemes of the UV region. 
As we shall  see in our analysis, the former two are very closely related. 
In the last scheme (PT) 
the UV scale dependence appears  naturally, in a form which - for two 
compact dimensions  case - agrees with the 
(heterotic) string. This  is supported by  findings in
\cite{Ghilencea:2002ff,Ghilencea:2003kt} where such a regularisation 
 recovered in a field theory approach,
the (limit of ``large'' radii of the) one loop string thresholds 
to the gauge couplings  in  4D N=1 toroidal orbifolds 
with  N=2 sub-sectors in the absence \cite{Ghilencea:2002ff} or presence 
\cite{Ghilencea:2003kt} of Wilson lines.

The plan of the paper is the following.
In the next section we review for one- and two-dimensional
orbifolds the   structure of 
the 4D KK  mass spectrum in the presence of non-zero Wilson 
lines vev's which ``commute'' with the orbifold projection 
of the model. The structure of the 4D
KK mass  spectrum is the starting point for the main 
analysis of this work (Section~\ref{msection}) where
we  compute  the  radiative corrections and
their dependence on the UV regulator/scale. 
The Appendix provides extensive and self-contained 
technical details for general series of Kaluza-Klein integrals 
that we encountered in  one-loop calculations,  in dimensional 
regularisation, zeta-function and proper-time cutoff regularisation. 
The  exact mathematical relation among these schemes is also provided. Such
results  can be useful for other applications involving one-loop radiative 
corrections from  compact dimensions.

\section{Orbifolds,  Wilson lines and 4D Kaluza-Klein mass spectrum.}


As an introduction  we review the effect of
Wilson lines on  the general form of the  4D Kaluza-Klein 
masses for one-  and two-dimensional field theory  orbifolds. 
Although some details  of the analysis may be different in specific
 models, the {\it   structure}  of the 4D  Kaluza-Klein masses that we find
in eqs.(\ref{massformula}), (\ref{mass3})  is general
\cite{Delgado:1998qr,Antoniadis:2001cv} and this  
is employed in  Section \ref{msection}. 

Consider  a one- and a two-dimensional orbifold of discrete group $Z_N$.  
For the one-dimensional case, its  action is 
$z\! \ra\! z' = \theta_l z$ and $z$ denotes the 
extra dimension. For two compact dimensions $z,\overline z$ one has
 $z \!\ra z'\! = \theta_l z$, ${\overline z}\!\ra\! 
{\overline z'}={\overline  \theta}_l\, 
{\overline z}$, with $\theta_l=\exp(2 i \pi\,l/N)$, $l=0,1,\cdots, N-1$. 
We denote  $\tilde\mu=\{\mu, z\}$ and 
$\tilde\mu=\{\mu, z,{\overline  z}\}$ for one and two compact dimensions
respectively, with $\mu=\overline{0,3}$. 
Then the gauge field $A_{\tilde\mu}$ and a  scalar multiplet
$\Phi$  in the fundamental representation transform 
as\footnote{There is an inconsistency in the notation in 
eq.(\ref{orbifold_cond}), (\ref{qmw}) and (\ref{ptheta}) in that 
for two compact dimensions
the fields $A_{\tilde \mu}$, $\Phi$ and operator $U$ are actually
functions of $(x,z,{\overline z})$ or  $(x,\theta_l z,{\overline \theta}_l
{\overline z})$ rather than  $(x,z)$ or  $(x,\theta_l z)$.}
\begin{eqnarray}\label{orbifold_cond}
A_{\tilde\mu}(x,\theta_l z) & = & \gamma_\theta \,
P_\theta\, A_{\tilde\mu} (x,z)\, P_\theta^{\dagger}, \qquad\,\, (x\in M^4) 
\nonumber\\
\Phi(x,\theta_l z) & = & P_\theta\, \Phi(x,z)
\end{eqnarray}
$\gamma_\theta=1$ for $\tilde \mu=\mu$, and
$\gamma_\theta=\theta_l^{-1}$ for the compact dimension(s) index. 
Conditions (\ref{orbifold_cond})  ensure that terms in the action
 as  $\vert D_{\tilde\mu} \Phi D^{\tilde\mu} \Phi\vert^2$ 
are invariant under the orbifold action.
Suppose the action has a symmetry $G^*$ 
before the orbifold action (\ref{orbifold_cond}),
so it is invariant under a gauge transformation~$U(x,z)$ 
\begin{eqnarray}\label{qmw}
A'_{\tilde\mu}(x,z) & = & U(x,z)\, A_{\tilde\mu}(x,z)\, U^\dagger (x,z)
- i\, U(x,z)\, \partial_{\tilde\mu}\, U^\dagger(x,z)
\nonumber\\
\Phi'(x,z) & = & U(x,z)\, \Phi(x,z)
\end{eqnarray}
Eq.(\ref{orbifold_cond}) is invariant 
under a gauge transformation $U(x,z)$ provided that
\begin{equation}\label{ptheta}
U(x,\theta_l z)\, P_\theta\, = \,P_\theta\, \, U(x,z)
\end{equation}
Eq.(\ref{ptheta})  gives the remaining gauge symmetry 
after imposing the orbifold condition (\ref{orbifold_cond}).
At fixed points $z_f=\theta_l z_f$, this is generated  by 
$G=\{T_a,\,\textrm{with}\,  T_a=P_\theta T_a P_\theta^\dagger\}$. 
For broken generators~($T_a^{*}$) with 
 $P_\theta T^{*}_a  P_\theta^\dagger=\omega^{k_a} T^{*}_a$,
($\omega\equiv e^{i 2\pi/N}$) and  with  $\omega^{k_a}=\theta_l$, 
the corresponding components  $A_z^a$ of the field $A_z$  respect
  $A_z^a(x,\theta z)=A_z^a(x,z)$, 
 and their non-zero vev's will break the group  $G$ further.

\subsection{One compact dimension: 
General structure of 4D Kaluza-Klein masses.}


The initial fields satisfy periodicity conditions with respect 
to the compact dimension $z$ 
\begin{eqnarray}\label{prd}
A_{\tilde\mu}(x,z+2 \pi R) & = & Q \, A_{\tilde\mu}(x,z) \, Q^{-1}
\nonumber\\
\Phi(x,z+2\pi R)& =&  Q\, \Phi(x,z)
\end{eqnarray}
where $Q$ is a global transformation. Eqs.(\ref{prd}) are
invariant under a gauge transformation $U(x,z)$ if
\begin{equation}\label{rwtypkq}
U(x,z+2\pi R)\, Q= Q\, U(x,z) 
\end{equation}
We now assume that $A_z$ of (\ref{orbifold_cond}) has some
non-zero  components in the Cartan-Weyl basis of G$^*$
(see discussion after eq.(\ref{ptheta})).
It is then easier to do calculations in a new gauge, 
with no  background field  
i.e. $A_z'=0$ which is  achieved by  a  $z$-dependent, 
non-periodic  gauge transformation. Then eq.(\ref{rwtypkq}) is not
respected and  eq.(\ref{prd}) will change for the gauge-transformed
(``primed'') fields. We consider  $A_z$  constant and for simplicity,
that it lies in the Cartan subalgebra of  G$^*$,
$A_z=A_z^I T_I^*$. The generators of the  group G satisfy: 
$[T_I, T_J]=0$; $[T_I,E_\alpha]=\alpha_I E_\alpha$; 
$I,J=1,\cdots,\textrm{rkG}$,  with $\alpha=1, \cdots,\textrm{dimG}
-{\textrm{rkG}}$. The non-periodic  gauge transformation is
\begin{eqnarray}
V(z)=e^{-i z A_z} Q^{-1}, \qquad (A_z= A_z^I \, T_I^*)
\end{eqnarray}
We use $A_{\mu}= A_{\mu}^I T_I+ A_{\mu}^\alpha
 E_\alpha$, $T_I \Phi_\lambda=\lambda_I \Phi_\lambda$ with
 $\Phi_\lambda$  the component $\lambda$ of the multiplet $\Phi$.
With~(\ref{qmw}) for $U=V$,
conditions (\ref{prd})  for the  fields transformed under 
 $V$ become
\begin{eqnarray}\label{symmetry}
A_\mu^{' I}(x,z+2\pi R) & = & A_\mu^{' I}(x,z), 
\qquad \quad \qquad A_z' =  0,
\nonumber\\
A_\mu^{' \alpha}(x,z+2\pi R) & = & e^{i \, 2  \pi \rho_\alpha} \,  
A_\mu^{' \alpha}(x, z),\qquad 
\rho_\alpha \equiv - R A_z^{ I} \alpha_I,\qquad 
\nonumber\\
\Phi_\lambda'(x, z+2 \pi R)  & = & e^{i\, 2 \pi \rho_\lambda}\,
\Phi_\lambda'(x,z), \qquad \,\,
\rho_\lambda \equiv - R A_z^{ I} \lambda_I
\end{eqnarray}
where $A_z$ respects  eq.(\ref{orbifold_cond}) and $\sigma=\alpha,
 \,(\lambda)$ for the adjoint (fundamental) representation.
In the following we  refer to $\rho_\sigma$ as  Wilson lines or 
``twist'' of higher dimensional fields with respect to the compact
dimension.  From Klein-Gordon equation with no gauge
background  (since $A_z'=0$) but with constraint (\ref{symmetry}),
we find  that  component fields with
``twist'' $\rho_\sigma$ ($\sigma=\alpha,\lambda$) have 4D modes
with  mass 
\begin{eqnarray}\label{massformula}
M_n^2(\sigma)=\chi^2+(n+\rho_{\sigma})^2\, \frac{1}{R^2} 
\end{eqnarray}
This provides the  structure of  4D Kaluza-Klein mass spectrum 
which takes account of  non-zero background fields $A_z^I$ or more
generally of  $\rho_\sigma$ twists in the ``new'' boundary
conditions eq.(\ref{symmetry}).
The contribution $\chi^2$ is only present if higher dimensional fields
such as $\Phi$ are massive\footnote{In such case $\chi$ will play the role
of infrared regulator in the radiative corrections to gauge couplings}.
For the gauge fields $\chi=0$ and $M_0(\alpha)\not =0$ if there is a
non-zero $\rho_\alpha$. As a result  the corresponding 
generator $E_\alpha$ is ``broken'' and the symmetry G is reduced.
See  \cite{Haba:2002py,Quiros:2003gg} for specific examples and 
related discussions.
Eq.(\ref{massformula}) will be used in Section~\ref{onedim_loop}. 

Although our derivation of the
mass formula (\ref{massformula}) is not necessarily general,
 the important point is that its structure  is generic
and appears in many orbifold compactifications $S_1/Z_2$, 
$S_1/Z_2\times Z_2$ \cite{Delgado:1998qr,Haba:2002py}  even 
 in the {\it absence} of 
 Wilson lines vev's $\rho_\sigma$. In  many cases   $\rho_\sigma$ 
is just  replaced  by a constant (``twist''), while its
value given in (\ref{symmetry}) is specific  to the case of Wilson  
line symmetry breaking only.

For generality the one-loop corrections from the KK modes are computed 
in Section \ref{onedim_loop} with  $\rho_\sigma$ an {\it arbitrary} parameter. 
Any model dependence will only involve minimal
redefinitions  of the parameters  $\rho_\sigma$, $R$ and $\chi$ of 
the model.

\subsection{Two compact dimensions: General structure of 
4D Kaluza-Klein masses.}


We repeat the  above analysis  for two compact dimensions.
For  compactifications on a (non-orthogonal) two-torus $T^2$,
the  higher dimensional  fields satisfy now  periodicity conditions 
with respect to shifts along both dimensions. Under the
following  shifts of $(z,\bar z)$  on the torus lattice:
$(z',\bar z')\equiv (z+2\pi R_2 \,e^{i\theta}, 
\bar z+2\pi R_2\,e^{-i\theta})$, 
$(z'',\bar z'')\equiv (z+2\pi R_1, \bar z+2\pi R_1)$, one has 
\begin{eqnarray}\label{periodic2D}
A_{\tilde\mu} (x; z',\bar z')=
Q \, A_{\tilde\mu} (x; z,\bar z)\, Q^\dagger,
& \qquad \qquad &\quad \,
\Phi (x; z', \bar z')=
Q \, \Phi (x; z,\bar z),
\nonumber\\
A_{\tilde\mu} (x; z'',\bar z'') = Q\, A_{\tilde\mu}(x;
z,\bar z)\, Q^\dagger, 
& \qquad \qquad & \quad
\Phi (x; z'', \bar z'') = Q\, \Phi(x; z,\bar z)
\end{eqnarray}
We  assume that $A_{z}, \,A_{\bar z}$ of (\ref{orbifold_cond}) 
have  non-zero  components in the Cartan-Weyl basis. For simplicity we 
take $A_{z}= A_{z}^I T^*_I$, $A_{\bar z}= A_{\bar z}^I T^*_I$ and
$A_z$, $A_{\bar z}$ constant. A  $z,\bar z$-dependent gauge 
transformation $V(z,\bar z)= \exp{(-i z
A_{z} - i \bar z A_{\bar z})} Q^{-1}$ gauges away the
constant gauge ``background'',  so   $A'_{z}=0$,
 $A'_{\bar  z}=0$.  After the  transformation $V$ the components   
in the Weyl-Cartan basis of the gauge-transformed  fields  satisfy
\begin{eqnarray}\label{twistedbc}
A_\mu^{' \alpha}(x; z',\bar z') =  e^{{ i \, 2 \pi \rho_{2,\alpha}}}
\,\,A_\mu^{'\alpha}(x; z,\bar z),
& \qquad &
\Phi_\lambda'(x; z',\bar z')
 = e^{i\, 2\pi \rho_{2,\lambda} }\, \, \Phi'_\lambda (x; z,\bar z),
\nonumber\\
A_\mu^{' \alpha}(x; z'',\bar z'')
 =  e^{ i\, 2 \pi \rho_{1,\alpha} } \,\,A_\mu^{'\alpha}(x; z, \bar z),
& \qquad &
\Phi'_\lambda (x; z'',\bar z'')  = 
e^{i\, 2 \pi \rho_{1,\lambda}} \,\, \Phi'_\lambda(x; z,\bar z),
\\
 \quad \rho_{1,\sigma}  
\equiv  - R_1 (A_z^I+A_{\bar z}^I)\,\sigma_I, &\qquad &
\rho_{2,\sigma} \equiv  - R_2 (A_{z}^I e^{i \theta}
+A_{\bar z}^I e^{-i\theta}) \,\sigma_I, 
\quad \sigma=\alpha,\lambda\label{vvv}
\end{eqnarray}
while $A_{\mu}^{' I}$ do not acquire any ``twist''.  Here  
$\sigma=\alpha$  ($\sigma=\!\lambda$) for adjoint  (fundamental)
representations respectively,  $\Phi_\lambda$ denotes 
a component $\lambda$ of the multiplet  $\Phi$ and we used 
$T_I\,\Phi_\lambda\,=\,\lambda_I\,\Phi_\lambda\, .$

From the Klein-Gordon equation with no gauge background\footnote{This
 was gauged away by $V$.} but with ``twisted'' boundary conditions
(\ref{twistedbc}) it can be shown  that the 
4D modes of component fields 
$A_\mu^{'\alpha}$, $\Phi'_\lambda$ acquire a mass \cite{Ghilencea:2003kt}
\begin{eqnarray}\label{mass3}
M^2_{n_1,n_2}(\sigma)& = & 
\frac{1}{\sin\theta^2 }\bigg\vert \, \frac{1}{R_2} (n_2+\rho_{2,\sigma})
- \frac{e^{i\theta}}{R_1} (n_1+\rho_{1,\sigma})\bigg\vert^2,
\quad \sigma=\alpha\,\,\, \textrm{or} \,\,\, \sigma=\lambda
\end{eqnarray}
or, in a different notation
\begin{eqnarray}\label{mass2}
M^2_{n_1,n_2}(\sigma)&=&
\frac{\mu^2}{T_2\, U_2} \, \vert n_2+\rho_{2,\sigma}-U
(n_1+\rho_{1,\sigma})\vert^2, \quad \sigma=\alpha,\lambda.
\end{eqnarray}
with 
\begin{eqnarray}
U\equiv U_1+i U_2& =& R_2/R_1\, e^{i\theta},\,\,\,\, (U_2>0);\, \qquad
T_2 (\mu) \equiv  \mu^2 R_1 R_2 \sin\theta
\end{eqnarray}
 $\theta$ is the angle between the two directions.
We introduced  $\mu$ a finite non-zero mass scale, to
ensure a dimensionless  definition for $T_2$; the dependence on $\mu$ 
 cancels out in $M_{n_1,n_2}$. Eqs.(\ref{twistedbc}) to (\ref{mass2}) 
show that  the symmetry G after orbifolding 
is further broken by the Wilson lines (\ref{vvv}) or  ``twist''  
$\rho_{i,\alpha}\not=0$ since then $M_{0,0}(\alpha)\not=0$,
the corresponding $A_\mu^{' \alpha}$  becomes massive 
and the generator $E_\alpha$ is ``broken''.

Eq.(\ref{mass3})  gives the  general structure of 4D KK masses for
$T^2$ with Wilson lines but also  for 
orbifolds\footnote{For example, for 
  $T^2/Z_2$ orbifold,  parity constraints of the orbifold
  impose  $\rho_{1,2}=0,1/2$ and the results for radiative
  corrections can in this case  be written in terms of those for
  $T^2$. (for $T^2/Z_N$,  $\rho_{1,2}=\exp(2 i
  \pi k/N)$, $k=0,1,..,N-1$).}
such as $T^2/Z_2$.
Additional constraints may  apply to  $A_z, A_{\bar z}$ and
thus to $\rho_{i,\sigma}$, $i=1,2$ which may take continuous/discrete 
values. However, for our analysis below
 we simply regard   $\rho_{i,\sigma}$ as {\it arbitrary,
  fixed} parameters. This allows  our  results  
to be applied to  specific models  (see examples in 
\cite{Antoniadis:2001cv})
with twisted boundary conditions, even in the  absence of 
 Wilson lines ($\rho_{i,\alpha}=0$).
   Model dependent constraints can  be implemented  
onto the   final results  
by using appropriate re-definitions of the parameters 
$\rho_i$, $U$, $T$.

\section{General form of  one-loop corrections.}
\label{msection}
\subsection{Case 1. One compact dimension.}
\label{onedim_loop}


Using the general structure of the KK mass spectrum in one and two
dimensional orbifolds with Wilson lines, eqs.(\ref{massformula}),
(\ref{mass2}), we  address the
implications for the radiative corrections to the 4D gauge couplings.
The one-loop correction  to the  gauge couplings
induced by the Kaluza-Klein states is given by the Coleman-Weinberg 
formula (see for example  \cite{Kaplunovsky:1987rp} for a general 
derivation of $\Omega_i(\sigma)$)
 \begin{figure}[t]
 \centerline{\psfig{figure=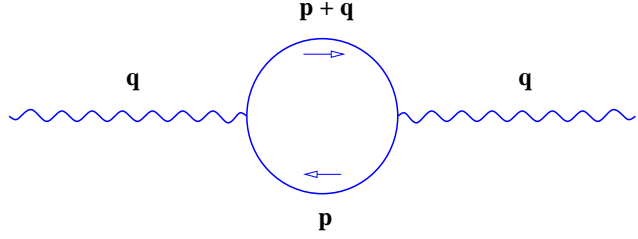,height=1.2in,width=3.3in,angle=0}}
 \caption{\small Generic one-loop diagram  contributing to $\Omega_i$,
  with KK modes in the loop.  Its value for   $q^2=0$  can be 
  read from eq.(\ref{eq0}),
  (\ref{twodim})  for  one and two compact dimensions. 
  See also Appendix A in \cite{dd10}.}
 \label{figure1}
 \end{figure}
\begin{equation}\label{eq0}
\frac{1}{\alpha_{i}}\bigg\vert_{{1-loop}}\!\!\!
=\frac{1}{\alpha_{i}}\bigg\vert_{{tree}}\!\!\!
    +\Omega_i^*,\,\quad 
\Omega_i^* =\sum_{r}\sum_{\sigma=\lambda,\alpha} \!\!
\Omega_i(\sigma), 
\,\quad
\Omega_i(\sigma)\equiv\! \frac{\beta_i(\sigma)}{4\pi} 
\sum_{m \in \bZ}' \int_{0}^{\infty}\!\frac{dt}{t} 
\, e^{-\pi\,t\, M^2_m(\sigma)/\mu^2}\bigg\vert_{\textrm{reg.}}
\end{equation}
where $\mu$ is a finite, non-zero mass parameter which enforces  
a dimensionless equation for $\Omega_i$.
We would like to mention  that the rhs  formula for $\Omega_i$ 
is obtained by  evaluating one loop diagrams for vanishing momentum
($q^2=0$), such as that of $\Pi(q^2)$  shown in  Figure  \ref{figure1}, 
with a tower of Kaluza-Klein (KK) states each of mass $M_m(\sigma)$ 
($m$ integer) present in the loop.  For more technical details on how 
to obtain this expression for $\Omega_i$  see for example  
Appendix A  in  \cite{dd10}, or \cite{IZ}. Note the
distinction  between the dependence (''running'') of the couplings on the 
momentum scale $q$ (see Figure~\ref{figure1})  for large $q$ and
given by $1/\alpha_i(q^2)-1/\alpha_i(0)=(\Pi_i(q^2)-\Pi_i(0))/\alpha_i(0)$,
and their dependence on the UV cutoff (regulator)
of   the theory that we compute in this work  for $q^2=0$ and  
given by $\Omega_i$. 
We will only briefly discuss the dependence  on
$q^2\!\not=\!0$ of the couplings, for  a detailed analysis  
see \cite{Goldberger:2002cz,toappear}.

$\Omega_i(\sigma)$ is thus the  contribution of an infinite 
tower of Kaluza-Klein modes 
associated with a state of charge  $\sigma$ in the Weyl-Cartan basis 
and  of mass ``shifted'' by $\rho (\sigma)$ real, with 
 $\sigma=\lambda,\alpha$ the 
weights/roots belonging to the representation $r$.
The ``primed'' sum over $m$ runs over all non-zero positive {\it and} 
negative
integers (levels). The case when this sum is  restricted to
positive (negative) levels only  will also be addressed. 
The effect of zero-modes is not included in $\Omega_i$ since 
their  presence is in general   model dependent. 
Thus their contribution should be added separately to $1/\alpha_i$.  
The important point to note is that
while the sums  over $r$ and  $\sigma=\alpha,\lambda$ in eq.(\ref{eq0})
depend on the field content and are thus model dependent,  the
integral and the sum in $\Omega_i$  over 
Kaluza-Klein modes of non-zero level depend only on the geometry of 
compactification. It is this  integral and  sum  over KK levels  which   
are difficult to perform exactly, and they are evaluated below.

Supersymmetry is not a necessary ingredient in formula
(\ref{eq0}). Supersymmetry is
however present in many models with compact dimensions which consider
 MSSM-like models as the viable ``low-energy'' limit.
Regarding  the beta functions $\beta_i$ one has (we suppress the 
subscript $i$) that  $\beta(\sigma)=k_r 
(\sigma_I \sigma^I)/{\textrm{rk} G}$ for $\sigma$ belonging  to 
representation $r$;  $k_r=\{-11/3,2/3,1/3\}$ for 
adjoint representations,  Weyl fermion and scalar respectively; $k_r$
essentially counts the degrees of freedom in the corresponding 
representations.
The Dynkin index $T(r)= (\sum_{\sigma} \sigma_I\sigma^I)_r/(\textrm{rk} G)$ 
where the sum  is over all weights/roots $\sigma$
belonging to representation $r$,  each occurring a  number 
of times equal to its multiplicity \cite{Slansky:yr}.  
With the definition $b_i(r)\equiv \sum_{\sigma}\beta_i(\sigma)$ 
for the weights $\sigma$ belonging to $r$ one has
$b_i\!=\!-11/3\, T_i(A)+2/3\,  T_i(R)+1/3\,  T_i(S)$, to
account for the adjoint, Weyl fermion in representation $R$ 
and scalar in representation $S$. In the supersymmetric case  
massive N=1 Kaluza-Klein states are organised as
N=2 hypermultiplets (vector supermultiplets) with  $b_i\!=\!2 T_i(R)$ 
($b_i\!=\!-2 T_i(A)$).

The subscript ``reg.'' shows that formula (\ref{eq0}) 
is not well defined in the UV region $t\ra 0$, and 
a UV regularisation is required. We assume $M_{m}(\sigma)\not=0$ so no
infrared (IR) regularisation (i.e. for $t\ra \infty$) 
is needed\footnote{See however the discussion in 
\cite{Ghilencea:2002ak} for the case of
two compact dimensions.}. The use of a particular 
regularisation is  in general dictated by the symmetries of
the initial, higher dimensional theory.  If a string
embedding exists for this theory, a proper-time cutoff (PT) 
regularisation (although not gauge invariant) is 
in  some cases the appropriate  choice
 \cite{Ghilencea:2002ff,Ghilencea:2003kt}. 
In the absence of a fully specified such theory and to keep the analysis 
general we compute $\Omega_i$ in three regularisation schemes: 
DR, ZR  and PT  regularisation.

\vspace{1cm}
\noindent
{\bf $\bullet$ Dimensional Regularisation (DR).}

\vspace{0.2cm}
\noindent
In this scheme $\Omega_i$ of 
eq.(\ref{eq0}) has  under the integral  $1/t$
replaced  by $1/t^{1+\epsilon}$ with $\epsilon\ra 0$  the UV
regulator. In such case $\mu$ is the arbitrary (finite, non-zero) 
mass scale introduced in the  DR scheme  in $d=4-\epsilon$ dimensions.
The evaluation of $\Omega_i$ in DR 
is rather long  and is presented in detail 
in  Appendix  \ref{computing_1}, eqs.(\ref{drint}) to
(\ref{drlast}).  The calculation  uses expansions
 in Hurwitz or Riemann zeta functions which do not necessarily
involve a Poisson re-summation of the ``original''  KK levels. This 
has the advantage that one  may  be able to identify 
which of the  {\it original} KK levels brings the leading contribution
to $\Omega_i$.
Using  eqs.(\ref{massformula}), (\ref{drint}), (\ref{dfdr}), (\ref{drlast})
one finds   for $\Omega_i$ in the DR scheme
\begin{eqnarray}\label{drresult}
\Omega_i{\bigg\vert_{DR}} & =& \!
\frac{\beta_i(\sigma)}{4\pi} 
\sum_{m \in \bZ}' \int_{0}^{\infty}\!\frac{dt}{t^{1+\epsilon}} 
\, e^{-\pi\,t\, M^2_m(\sigma)/\mu^2}
\nonumber\\
\nonumber\\
& = &
\frac{\beta_i(\sigma)}{4\pi}\,
\bigg\{\,\frac{1}{\epsilon}
-\ln\frac{(R\mu)^2}{\pi e^\gamma} 
- 
\ln \left\vert\,\frac{ 2\, \sin
\pi({\rho_{\sigma}}+i \chi R)}{\rho_\sigma+ i \chi R }\right\vert^2\bigg\}
\end{eqnarray}
The presence of the pole in $\epsilon$ accounts for an UV divergence. 
To find the {\it scale} dependence of this divergence in
DR  one may in general  introduce  a small/infrared mass shift 
$\chi$ of the 
momentum of the KK  state. One would then  expect the emergence in the 
final result of
a term $\chi/\epsilon$ to account for a linear divergence (in scale), 
given the
extra dimension present. However this procedure does not apply to the
case with one compact dimension only\footnote{This  is somewhat
similar to computing  $\int d^4 p/p^2$ which in cutoff
regularisation is quadratically divergent while in DR is vanishing. 
However, a small mass shift $\chi$ of the momentum leads to  
$\int d^4 p/(p^2+\chi^2)$ which has a pole in DR, which signals the
usual quadratic divergence. In our case, even adding a small 
(mass)$^2$ shift (accounted  for by $\chi^2$) does not introduce a 
scale dependence of the divergence in (\ref{drresult}), such 
as $\chi/\epsilon$. 
For two (even number of)
compact dimensions, this procedure in DR does lead 
to the scale  dependence of the leading divergence, as opposed to
 the case of one (odd number of) extra dimension(s). See also
eq.(\ref{prtq}) in the Appendix which shows the emergence of a
linear divergence in DR when summing over {\it positive (negative) modes
 only}.}. Therefore, unlike the case of two compact dimensions to be
discussed later, the presence of the  pole alone  does not tell us 
the nature of the scale dependence of the UV divergence.  
Note also that a single state
(such as the zero mode for example) gives a leading one-loop 
contribution 
proportional to $-1/\epsilon$, which is of the same form 
but of {\it opposite 
sign} to that found in eq.(\ref{drresult}) 
for the whole Kaluza-Klein tower, excluding the zero-mode~! 
(compare $\cR_\epsilon$ vs. {\it finite} $\cR_\epsilon^T$ 
in eq.(\ref{drlast}) of Appendix \ref{computing_1}). 
Further, this $-1/\epsilon$ pole due to a single  
state is also known to correspond  in 4D to a UV divergence only
logarithmic  in scale  (rather than  linear).  
Note however that the  change of the couplings with  momentum
$q$ in Fig.\ref{figure1} is indeed linear in the momentum scale~$q$
  and dominates if $(q R)^2\! \geq  \! \cO(1)$
  \cite{Goldberger:2002cz,toappear}.

If $\rho_\sigma$ is a non-zero integer, there exists a level $n_0$ 
such as $M_{n_0}= \chi$ and then $\chi$ plays the 
role of  an IR regulator in eqs.(\ref{eq0}), (\ref{drresult}), 
and  ensures that the term
$\ln[\sin\pi(\rho_\sigma+i\chi R)]$  remains finite. If $\rho_\sigma$,
$\chi$ vanish the last term in
(\ref{drresult}) vanishes and one is left with the correction in the
absence of the ``twist'' or Wilson line background~$\rho_\sigma$.

In deriving $\Omega_i$ we summed over both positive and
negative Kaluza-Klein levels, as shown in eq.(\ref{eq0}). However, it 
is useful  (particularly for  orbifolds case)
to analyse the effect of summing separately
the contributions of the positive (negative) levels. 
In such a case the corresponding value of $\Omega_i$, denoted $\Omega_i^+$
($\Omega_i^-$),  is  computed in a  similar way.  
The  result,  derived  in  Appendix  \ref{computing_1}, 
eqs.(\ref{pmmm}), (\ref{prtq}) is
\vspace{0.17cm}
\begin{eqnarray}
\Omega_i^\pm\bigg\vert_{DR} 
&=&
\frac{\beta_i(\sigma)}{4\pi}\,
\bigg\{
\frac{1}{2\epsilon} \pm \frac{\rho_\sigma}{\epsilon}+
\ln\Big\vert\Gamma(1 \pm \rho_\sigma+i \chi R)\Big\vert^2
-\ln(2\pi)-\Big[\frac{1}{2} \pm \rho_{\sigma}\Big]\ln \frac{(R\mu)^2}{\pi
  e^\gamma}
\bigg\}
\end{eqnarray}
The divergent  terms of $\Omega_i^\pm$ are then
\begin{equation}\label{smr2}
\Omega_i^\pm\bigg\vert_{DR}\sim 1/(2\,\epsilon) \pm
\rho_\sigma/\epsilon
\end{equation}
The presence of the  additional divergence $\rho_\sigma/\epsilon$ 
is triggered by a  non-zero  background/twist $\rho_\sigma$, and 
is cancelled  in the  sum $\Omega_i^+ +\Omega_i^-$ 
of both positive and negative Kaluza-Klein
levels, giving  the overall  result $\Omega_i$ 
in (\ref{drresult}). If $\rho_\sigma$ has
the value given in (\ref{symmetry}) and is thus proportional to the
vev of $A_z$ and to $R$, then   $\rho_\sigma/\epsilon$ may be regarded as a 
divergence linear in scale. It is also possible 
that in some models one may actually have $\rho_\sigma$ a {\it constant},
for example $\rho_\sigma=+ 1/2$,  
(or $-1/2$) then $\Omega_i^-$ ($\Omega_i^+$) are  {\it finite}
respectively, and the {\it overall} divergence in 
$\Omega_i=\Omega_i^+ +\Omega_i^-$  comes entirely from 
$\Omega_i^+$ ($\Omega_i^-$) respectively! 
To conclude, the positive and negative Kaluza-Klein 
levels propagating in opposite directions 
in the compact dimension, with a non-zero background/twist $\rho_\sigma$,
contribute by different amounts to the overall divergence of
$\Omega_i$; in special cases the positive 
or negative levels alone  give (one-loop) {\it finite} contributions !
 This concludes  our analysis in the DR scheme.

\vspace{0.68cm}
\noindent
{\bf $\bullet$ $\zeta$-function Regularisation (ZR).}

\vspace{0.2cm}
\noindent
Alternatively, one can employ a $\zeta$-function regularisation
of  $\Omega_i$.
In this case the correction is given  (up to a factor
$\beta_i(\sigma)/(4\pi)$),  by  the derivative of 
$\zeta$-function associated with the Laplacian, evaluated in origin. 
As detailed in  Appendix \ref{computing_zeta} 
this  means that $\Omega_i$ in this scheme is just the 
derivative with respect to $\epsilon$ of the value obtained in 
the DR scheme (divided by $\Gamma[-\epsilon]$), and evaluated 
for $\epsilon=0$. From  eqs.(\ref{massformula}), (\ref{eq0}), 
(\ref{zetafl}), (\ref{rtpgl})  one obtains the
value of $\Omega_i$  in the  ZR scheme 
\begin{eqnarray}\label{zeta_result}
\Omega_i\bigg\vert_{ZR} & = & \frac{\beta_i(\sigma)}{4\pi} 
\frac{d}{d\epsilon}  
\bigg\{
\frac{-\pi^{-\epsilon}}{\Gamma[-\epsilon]}
\sum_{m \in \bZ}' \int_{0}^{\infty}\!\frac{dt}{t^{1+\epsilon}} \, 
e^{-\pi\,t\, M^2_m(\sigma)/\mu^2}\bigg\}\bigg\vert_{\epsilon\ra 0}
\nonumber\\
\nonumber\\
&=&
\frac{\beta_i(\sigma)}{4\pi}\,
\bigg\{
-\ln (R\mu)^2
- 
\ln\left\vert\,\frac{ 2\, \sin
\pi({\rho_{\sigma}}+i \chi R)}
{\rho_\sigma+ i \chi R}\right\vert^2\,
\bigg\}
\end{eqnarray}
This result is similar to that found  in the DR scheme, with the
notable  difference that there is no pole structure present.
The above result is only logarithmically dependent on the mass scale $\mu$. 
As discussed in Appendix \ref{computing_zeta}, $\mu$ 
plays in the case of $\zeta$-function regularisation
the role of  the UV cutoff of the model.
Finally, note that the contribution  of a zero-mode  - if included -
would bring a similar dependence on  $\mu$ but of {\it opposite} sign
to cancel 
this dependence in the total sum (see also $\cR_\zeta$ and  
$\cR_\zeta^T$ in Appendix~\ref{computing_zeta}).

One can show that the separate contributions to $\Omega_i$ of positive and 
negative Kaluza-Klein modes are different due to the
asymmetry introduced by the  Wilson lines or twist $\rho_\sigma$. 
The results are
denoted $\Omega_i^+$ ($\Omega_i^-$) respectively, and  
are given by  eq.(\ref{zrzr}) 
\begin{eqnarray}
\Omega_i^\pm \bigg\vert_{ZR} 
& = & \frac{\beta_i(\sigma)}{4\pi}\bigg\{
\ln\Big\vert \Gamma[1 \pm \rho_\sigma+i \chi R ]\Big\vert^2
-\ln(2\pi)-\Big[\frac12 \pm \rho_\sigma \Big]\ln{(R\mu)^2} \bigg\}
\end{eqnarray}
so the  positive (negative) modes again bring  a different UV behaviour
($\mu$ dependence)
\begin{equation}\label{smr1}
\Omega_i^\pm\bigg\vert_{ZR} \sim -(1/2 \pm  \rho_\sigma) \ln (R\mu)^2
\end{equation}
For $\rho_\sigma$ just a {\it constant}, the $\rho_\sigma$-dependent 
term is just an additional logarithmic correction 
(in $\mu$ or $R$) to the couplings. 
However, in the case
 $\rho_\sigma$ is indeed due to a non-zero   Wilson line vev 
(from  initial $A_z$ gauge fields), a {\it linear} dependence of 
the couplings on this vev/scale emerges. This term can then  
have significant
implications for the value of the gauge couplings.  As it was
 the case in
the  DR  scheme, such terms cancel in the sum of positive and 
negative modes' contributions.
A special case is  $\rho_{\sigma}=\mp 1/2$ when the coefficient of the 
logarithmic UV divergence (in $\mu$)  of  $\Omega_i^{\pm}$
is vanishing, and $\Omega_i^+\, (\Omega_i^-)$  has no $\mu$
dependence, with similarities to the DR case.

\vspace{1cm}
\noindent
{\bf $\bullet$ Proper-time Regularisation (PT)}.

\vspace{0.25cm}
\noindent
The above  results for $\Omega_i$ can be compared to that obtained in the
proper-time regularisation.  In this
regularisation $\Omega_i$ of eq.(\ref{eq0}) 
has the lower limit of its integral set equal to $\xi>0$, 
where $\xi\!\ra\! 0$ is a dimensionless UV regulator. 
For details of the calculation of $\Omega_i$  in this
scheme see Appendix~\ref{computing_2}  and ref.
\cite{Ghilencea:2003kt} (Appendix  A-1). From
eqs.(\ref{massformula}), (\ref{prtw}), (\ref{ptpt}), (\ref{ptclast}) 
and with
the notation  $\Lambda^2\equiv \mu^2/\xi$ one obtains for 
$\Omega_i$  in the PT scheme
\begin{eqnarray}\label{ptc_res}
\Omega_i\bigg\vert_{PT} &=&
\frac{\beta_i(\sigma)}{4\pi} 
\sum_{m \in \bZ}' \int_{\xi}^{\infty}\!\frac{dt}{t} 
\, e^{-\pi\,t\, M^2_m(\sigma)/\mu^2}
\nonumber\\
\nonumber\\
&=& \frac{\beta_i(\sigma)}{4\pi}\,
\bigg\{2 R \Lambda - \ln \frac{(R \Lambda)^2}{\pi
    e^\gamma}
- \ln\left\vert\,\frac{ 2\, \sin
\pi({\rho_\sigma}\!+\! i \chi R)}
{\rho_\sigma+  i \chi R} \right\vert^2
\bigg\}
\end{eqnarray}
The $\xi$ dependent terms combine naturally 
with the scale $\mu$  to define  the UV cut-off $\Lambda$ of the model 
 and one obtains a dependence on  $\Lambda R$
only.  Unlike the DR and ZR cases, a zero-mode 
contribution to the above result - if included -
would {\it not} cancel the leading linear divergence (in $\Lambda\sim
1/\sqrt\xi$),  but
only the logarithmic one (for more details compare $\cR_\xi$ and
$\cR_\xi^T$ in Appendix \ref{computing_2}, eq.(\ref{ptclast})).

What is the meaning  of the individual contributions to
$\Omega_i$ ? Technical details show that the term $\ln\vert
\rho_\sigma+i\chi R\vert$
is similar  to a contribution  corresponding to a massive Kaluza-Klein
state of level zero. It may be interpreted  as a one-loop effect of
this state between the compactification scale $1/R$ and the scale
set by  the Wilson lines vev's, $\sigma_I \!<\! A_z^I\!>\!$ with $\sigma$
accounting for a root/weight. The term  $\ln[\sin(..)]$ in
(\ref{ptc_res}) is an effect due to ``Poisson re-summed'' (PR)
Kaluza-Klein states (see eq.(\ref{p_resumation})), with the dominant 
contribution from the lower PR  levels.  Further, 
the logarithm $\ln(\Lambda R)$ can be thought of as a one loop effect
from the compactification scale to the UV cutoff scale $\Lambda$. 
Finally,  the term $\Lambda R$ is  due to the presence of a 
{\it  large enough} number of Kaluza-Klein modes  which 
enable the Poisson re-summation. This term 
is due to  the  Poisson re-summed mode of zero-level.  Thus one 
should expect
$\Lambda R\gg 1$ because $\Lambda R$ approximates the number of
Kaluza-Klein modes.  In fact the  PT result (\ref{ptc_res}) is valid
provided that 
\begin{eqnarray}
\max \left\{ {1}/{R^2}, \, \,\chi^2,\,\,
  (<\!A_z^I\!>\!\sigma_I)^2 +\chi^2\right\}\ll\Lambda^2
\end{eqnarray}
derived from eq.(\ref{conditions}) of Appendix \ref{computing_2}. Here
we replaced $\rho_\sigma$ in terms of the  vev's of $A_z^I$ as in
eq.(\ref{symmetry}). More generally,
for arbitrary  $\rho_\sigma$ this condition is 
\begin{equation}
\max\left\{ {1}/{R^2}, \,\,
\chi^2,\,\,{\rho_\sigma^2}/{R^2}+\chi^2\right\}\ll \Lambda^2
\end{equation}
Therefore the result in the PT scheme is valid if $R$ is large 
 (in UV cutoff units) and if 
the gauge symmetry breaking  vev's or $(\rho_\sigma/R)^2$
and the mass scale  $\chi^2$ have a sum much smaller than the UV
cutoff.  Note that these constraints 
are not shared  by the   DR or ZR counterparts computed above. 
This is important for in
general to avoid a large UV sensitivity of the couplings 
one would like to have 
$\Lambda R\approx 1$ which is a region for which the PT result 
does not hold accurately.   From comparing it with its DR counterpart, 
the presence of the pole $1/\epsilon$ of the latter may indicate that
even if $\Lambda R$ is made smaller, of order unity,
a UV divergence is still manifest. Finally, if one considered a
string embedding of these models, the string counterpart
of $\Lambda R\approx 1$ would be $M_s R\approx 1$  with $M_s$  the string
scale. In this case  string effects  due to
additional  (winding) states not present in field theory may become important.

Comparing the three results for $\Omega_i$  obtained in these 
different regularisation schemes one observes that the finite
(regulator independent) part is the same in all regularisations. 
This is a strong consistency  check of the calculation.  
Regarding the divergent (i.e. regulator dependent) part, note that 
the $1/\epsilon$ term of 
DR is replaced in the PT cut-off regularisation  by the  
$\xi$  ($\Lambda$) dependent  term, accounting  
 for a  linear  divergence. Note  that 
the ZR  counterpart   has only (rather ``mild'') a logarithmic 
UV divergence. 
Eqs.(\ref{drresult}) to (\ref{ptc_res}) generalise
the results  \cite{dd10} for one compact dimension,  in the presence
of Wilson lines/twists $\rho_\sigma$.

\vspace{0.5cm}
\subsection{Case 2. Two compact dimensions.}
\label{two_dim_loop}


We consider now the case of a  two dimensional compactification
on\footnote{With appropriate replacements $\rho_{1,2}\ra 0,1/2$,
the results below can also be extended to  the $T^2/Z_2$ orbifold.}
 $T^2$ with Wilson lines. With   the structure of the mass spectrum of
eq.(\ref{mass2}) we again compute the general  form  of the  
correction to  the 4D gauge couplings due to non-zero level 
Kaluza-Klein modes in the presence of Wilson lines.
This correction can be applied to a large class of models 
\cite{Antoniadis:2001cv}. Formally, the correction  is 
\begin{equation}\label{twodim}
\Omega_i^*=\sum_{r}\sum_{\sigma=\lambda,\alpha} 
\Omega_i(\sigma), 
\qquad\quad
\Omega_i(\sigma)\equiv \frac{\beta_i(\sigma)}{4\pi} 
\sum_{n_1,n_2 \in \bZ}'
\int_{0}^{\infty}\frac{dt}{t} 
\, e^{-\pi\,t\, M^2_{n_1,n_2}(\sigma)/\mu^2}\bigg\vert_{\textrm{reg.}}
\end{equation}
Similarly to the case of  one extra dimension, $\Omega_i$ is obtained
by computing  one loop diagrams evaluated for $q^2=0$
(Figure \ref{figure1}) 
with Kaluza-Klein states of mass $M_{n_1,n_2}(\sigma)$
in the loop.

In the following we perform - for $\sigma$ fixed - the integral
and the sums over $(n_1,n_2)\not=(0,0)$ in  eq.(\ref{twodim}).
Any model dependence (beta functions $\beta_i(\sigma)$, sums over 
weights $\sigma$,
representations $r$) can then easily be implemented on the final
result for $\Omega_i^*$. The  presence  of the state $(n_1,n_2)=(0,0)$ 
is  model dependent and its contribution should be considered
separately.  We again discuss the value of $\Omega_i$ in DR, ZR and PT
regularisation schemes 
for the UV divergence ($t\ra 0$) of eq.(\ref{twodim}). 
We assume $M_{n_1,n_2}\not=0$ for all integers, so no 
IR divergence (at $t\ra \infty$)  exists. However, if 
there exists a pair $(n_1,n_2)$ for 
which $M_{n_1,n_2}=0$ see the results in the PT scheme of
\cite{Ghilencea:2002ak} and the discussion in the DR scheme to follow.

\vspace{0.6cm}
\noindent
{\bf $\bullet$ Dimensional Regularisation (DR).}

\vspace{0.2cm}
\noindent
In the DR scheme $\Omega_i$ is defined 
with  $1/t$ under its integral replaced by $1/t^{1+\epsilon}$ where 
$\epsilon\ra 0$ is the UV regulator. The calculation is rather technical
and is presented in  Appendix \ref{appendixC},  eq.(\ref{ldef}) to
(\ref{drl}), where the sums over $n_{1,2}$ and  integral in
(\ref{twodim}) are evaluated. Using 
eqs.(\ref{mass2}), (\ref{twodim}), (\ref{eqrt}), (\ref{drl}) and 
(\ref{idt}), one obtains $\Omega_i$ in the DR scheme 
\begin{eqnarray} \label{drl2}
 \Omega_i\bigg\vert_{DR} 
& = &
\frac{\beta_i(\sigma)}{4\pi} 
\sum_{n_1,n_2 \in \bZ}'
\int_{0}^{\infty}\frac{dt}{t^{1+\epsilon}} 
\, e^{-\pi\,t\, M^2_{n_1,n_2}(\sigma)/\mu^2}
\nonumber\\
\nonumber\\
& = & 
\frac{\beta_i(\sigma)}{4\pi}\bigg\{
 \frac{1}{\epsilon} -
\ln\frac{T_2 U_2}{\pi e^\gamma} -\ln\bigg\vert 
\frac{
\vartheta_1({\rho_{2,\sigma}}-
U {\rho_{1,\sigma}}\vert U)}{ ( \rho_{2,\sigma}- 
U \rho_{1,\sigma})\,\, \eta(U)} \, e^{i \pi U \rho^2_{1,\sigma}}
\bigg\vert^2 
\bigg\}
\end{eqnarray}
The special functions $\eta,\,\vartheta_1$ 
are defined in Appendix \ref{appendixG}. 
The  pole $1/\epsilon$ accounts for divergences up to  quadratic
level. How can we see this? By introducing a small (mass)$^2$ shift
$\mu^2 \delta$  to $M_{n_1,n_2}^2$, ($\delta$  dimensionless,
$\delta\ll 1$), 
i.e. $M_{n_1,n_2}^2\ra M_{n_1,n_2}^2+ \mu^2\delta$
under the integral in (\ref{twodim})  and computing 
the integral in this more general case  one obtains for $\Omega_i$, 
in addition to the divergence  $1/\epsilon$, a contribution  
$\pi \delta T_2/\epsilon$. This  is a quadratic 
divergence in scale ($T_2$ ``contains'' a $\mu^2$)
that $1/\epsilon$ term  effectively signals  in
eq.(\ref{drl2}) and (\ref{drl}). For additional technical details  see 
Appendix~\ref{appendixC}, eq.(\ref{l1l2l3}), (\ref{lpp})\footnote{This also
has consequences for  the change of the gauge
couplings with  momentum $q$ in 
Fig.\ref{figure1} as discussed in \cite{toappear}.}.  
The emergence of the additional
scale dependent  contribution $\pi \delta T_2/\epsilon$ 
is  to be contrasted  with what happened 
in DR in the  one extra dimension case  already discussed, where a small
mass shift did not introduce  a scale dependence of 
the UV divergence. 
This is due to the different UV behaviour of models with 
one (or odd  number of) and two  (or even number of) compact
dimensions, respectively. Note that in the special case when
there exists a pair $(n_1,n_2)$ such as  $M_{n_1,n_2}=0$, an
IR regulator  - in addition to the UV one  -
is required in eq.(\ref{twodim}), (\ref{drl2}) to ensure the convergence
of the integral at $t\ra \infty$.
 The aforementioned shift $\mu^2 \delta$ of the
KK masses would in such special case  
 act as an IR regulator in (\ref{twodim}) and one would
obtain in (\ref{drl2}) 
a term $\pi\delta T_2/\epsilon$ which  represents an IR-UV 
``mixing'' term 
between the IR sector ($\delta$) and UV sector ($\epsilon$) of the theory.
For a discussion on this UV-IR mixing 
see \cite{Ghilencea:2002ak,toappear} where 
its string theory interpretation is also presented.
Finally, considerations similar to those for one extra dimension 
apply for the separate  
role of negative or positive Kaluza-Klein levels, respectively.

\vspace{0.5cm}
\noindent
{\bf $\bullet$ $\zeta$-function Regularisation (ZR).}

\vspace{0.2cm}
\noindent
In this scheme $\Omega_i$ is  related   
to the derivative of the Zeta-function associated with the
Laplacian, as discussed in Appendix \ref{computing_zeta_two}. In fact 
$\Omega_i$ in ZR is the derivative with respect to $\epsilon$ of 
$\Omega_i$ in DR divided by $\Gamma[-\epsilon]$, and evaluated for 
$\epsilon=0$. Using eqs.(\ref{mass2}), (\ref{zzrr}), (\ref{thrw}),
(\ref{idt})  one finds  $\Omega_i$ in the  ZR scheme
\begin{eqnarray} \label{zeta_reg2}
 \Omega_i\bigg\vert_{ZR}
&=&
\frac{\beta_i(\sigma)}{4\pi} 
\frac{d}{d\epsilon} \bigg\{ 
\frac{-\pi^{-\epsilon}}{\Gamma[-\epsilon]}
\sum_{n_1,n_2 \in \bZ}' \int_{0}^{\infty}\frac{dt}{t^{1+\epsilon}} 
\, e^{-\pi\,t\, M^2_{n_1,n_2}(\sigma)/\mu^2}\bigg\}
\bigg\vert_{\epsilon\ra 0}
\nonumber\\
\nonumber\\
&=& 
\frac{\beta_i(\sigma)}{4\pi}\bigg\{
-
\ln [T_2 U_2 ] -\ln\bigg\vert 
\frac{\vartheta_1({\rho_{2,\sigma}}-
U {\rho_{1,\sigma}}\vert U)}{  ( \rho_{2,\sigma}- 
U \rho_{1,\sigma})\,\, \eta(U)} \, e^{i \pi U \rho^2_{1,\sigma}}
\bigg\vert^2 
\bigg\}
\end{eqnarray}
This result has a form similar to that in the DR scheme 
from which the  pole structure has been subtracted.
The $\mu$ scale dependence ``hidden'' in $T_2$  should in this case be
regarded as the UV cutoff as discussed in Appendix
\ref{computing_zeta_two}, eq.(\ref{wzprtghkjl}).
In this scheme there is thus only a logarithmic
dependence on the UV cutoff.  Finally, the finite part is 
similar to that obtained in the DR scheme. 
It would be of phenomenological interest 
to know which higher dimensional theories would require  such a 
regularisation, since in this case the UV cut-off dependence of 
the couplings is milder and the models would then  have less amount 
of sensitivity to this cut-off scale, possibly  similar to that of 
MSSM-like models. 

\vspace{0.6cm}
\noindent
{\bf $\bullet$ Proper-time Regularisation (PT).}

\vspace{0.3cm}
\noindent
Finally, for a comparison we include here the value of
$\Omega_i$ in the proper-time cutoff regularisation scheme 
(PT) \cite{Ghilencea:2003kt}.  In this scheme $\Omega_i$ of
(\ref{twodim}) is defined with a (dimensionless) cutoff $\xi\!\ra\! 0$
in the lower limit of its integral  which acts as an UV regulator.
After a long calculation one  obtains the result (for details see
eqs.(\ref{twodim}), (\ref{lpwtr}), (\ref{ptcreg}), (\ref{idt}) and also 
eq.(52) in \cite{Ghilencea:2003kt})
\begin{eqnarray}\label{ptc_result}
\!\!\!\Omega_i\bigg\vert_{PT}
&=&
\frac{\beta_i(\sigma)}{4\pi} 
\sum_{n_1,n_2 \in \bZ}'
\int_{\xi}^{\infty}\frac{dt}{t} 
\, e^{-\pi\,t\, M^2_{n_1,n_2}(\sigma)/\mu^2}
\nonumber\\
\nonumber\\
&=&
\frac{\beta_i(\sigma)}{4\pi}
\bigg\{ \frac{T_2}{\xi}
- \ln \frac{[(T_2/\xi)\, U_2]}{\pi e^\gamma}- \ln\bigg\vert 
\frac{\vartheta_1({\rho_{2,\sigma}}-U {\rho_{1,\sigma}}\vert
U)}{ (\rho_{2,\sigma}-U \rho_{1,\sigma})\, \eta(U)}
\, e^{i \pi U \rho^2_{1,\sigma}}
\bigg\vert^2
\bigg\}
\end{eqnarray}
Eq.(\ref{ptc_result})  is valid if (see  eq.(\ref{vevs}) and
definition (\ref{vvv}))
\begin{equation}\label{cprtlgm}
\max \left\{\frac{1}{R_1}, \, \, \frac{1}{R_2 \sin\theta}, 
\,\,  <\! A_z^I\!>\! \sigma_I, \,\,  <\!A_{\bar z}^I \!>\!\sigma_I\right\}
\ll \Lambda,\qquad \Lambda^2\equiv \frac{\mu^2}{\xi}
\end{equation}
This condition requires ``large'' compactification radii 
(in UV cutoff units)  and symmetry breaking vev's much smaller 
than $\Lambda$. Here we replaced $\rho_{i,\sigma}$ in terms of 
the vev's of $A_z^I$, eq.(\ref{vvv}) but  for 
arbitrary $\rho_{i,\sigma}$ this condition is:
 $\max\{1/R_1,1/R_2 \sin\theta,
\vert\rho_{1,\sigma}\vert/R_1, \vert\rho_{2,\sigma}- U
\rho_{1,\sigma}\vert/(R_2\sin\theta)\}\ll \Lambda$.

Eq.(\ref{ptc_result})  shows the presence of a UV quadratic 
divergent term  also known as ``power-like'' threshold, given by 
$T_2/\xi=\Lambda^2 R_1 R_2 \sin\theta$
where $\Lambda^2\sim 1/\xi$ is the UV cutoff scale. A logarithmic 
correction
is also present, $\ln (T_2/\xi)=\ln (\Lambda R_1 R_2 \sin\theta)$, as
well as a $\ln U_2=\ln (R_2 \sin\theta/R_1)$ part. The remaining 
term in $\Omega_i$ includes effects due to 
non-zero  $\rho_\sigma$ which bring in a finite,
regulator independent correction.

\vspace{0.5cm}

The field theory result (\ref{ptc_result}) has a great advantage over
its DR and ZR counterparts in that it allows a straightforward
comparison with the heterotic string result with Wilson lines. 
Here we refer to the 4D N=1 heterotic string orbifolds with
N=2 sectors (of unrotated $T^2$) and Wilson lines \cite{Mayr:1995rx}
when this string result is considered 
in the limit of large compactification radii/area (in string units)
\cite{Ghilencea:2003kt}, as required by eq.(\ref{cprtlgm}). 
The UV regulator 
$\xi\sim 1/\Lambda^2$ has a  natural counterpart in  the (heterotic) 
string in  $\alpha'\sim 1/M_s^2$ ($M_s$ is the string scale).
Therefore,   $T_2/\xi$ of (\ref{ptc_result})
has a counterpart at the string level in $T_2/\alpha'$, where
$T_2/\alpha'$ is the  (imaginary part of the)  K\"ahler structure
moduli. With the correspondence of the fundamental
lengths in field and string theory respectively, $\xi\leftrightarrow
\alpha'$, the result 
(\ref{ptc_result}) is indeed similar \cite{Ghilencea:2003kt}
 to the  limit of large radii of the heterotic string result 
\cite{Mayr:1995rx}.
Such  agreement provides support for this regularisation scheme
in the field theory approach, although it is not gauge invariant. 
String theory also brings additional corrections,
non-perturbative on the field theory side (world-sheet instantons)
but their effect is exponentially suppressed  $\cO(e^{-T_2/\alpha'})$
\cite{Mayr:1995rx}. For more details on the exact  link with the
corrections to the gauge couplings due to   the heterotic string  
with Wilson lines present,   see \cite{Ghilencea:2003kt}.

The effective field theory 
result (\ref{ptc_result}) has an interesting limit, that of
vanishing Wilson lines vev's or ``twists'' $\rho_{i,\sigma}$. 
For $\rho_{i,\sigma}\!\ra\! 0$ ($\sigma$ fixed) after using the 
relations in  eq.(\ref{ddt}), (\ref{th}) one finds 
\begin{equation}\label{pwl}
\Omega_i\bigg\vert_{PT}(\rho_{i,\sigma}\!\ra\! 0)
=- \frac{\beta_i(\sigma)}{4\pi} 
\ln\Big[4 \pi e^{-\gamma} \, e^{-{T_2}/{\xi}} \,\, {({T_2}/{\xi})} \,\, U_2
\, \vert \eta(U)\vert^4\Big],\qquad {T_2}/{\xi} = \Lambda^2 R_1
R_2 \sin\theta 
\end{equation}
For  two compact dimensions this result generalises
the  ``power-law'' corrections (in the UV cutoff) of ref.\cite{dd10}, 
by including the  dependence on $U=R_2/R_1 e^{i\theta}$.

The field theory result (\ref{pwl}) is itself  the exact limit
\cite{Ghilencea:2002ff,Ghilencea:2002ak} 
of ``large $R_{1,2}$'' (in string units) of a similar result in 
4D N=1 heterotic string orbifolds with N=2 sectors (of unrotated
$T^2$) without  Wilson lines \cite{Dixon:1990pc}.  
The only difference\footnote{See however
ref.\cite{Ghilencea:2002ak} and the discussion 
in the DR scheme.} between $\Omega_i$ of (\ref{pwl})
and  the above  limit of the string result \cite{Dixon:1990pc} is 
that the leading term $T_2/\xi$ in $\Omega_i$
has a coefficient which depends on the regulator choice 
($\xi$) while in string case at    
``large $R_{1,2}$'' the leading term  is\footnote{The presence of
 $\pi/3$ is a ``remnant'' of the modular invariance symmetry 
of the string.}
   $(\pi/3) T_2/\alpha'$. 
With the correspondence $\xi\leftrightarrow \alpha'$ mentioned before,
the exact matching of these two terms thus requires a re-definition 
of the  PT
regulator  $\xi\ra (3/\pi)\, \xi$ or equivalently $\Lambda^2\ra \pi/3\,
\Lambda^2$.  Such specific normalisation of $\xi$  (or $\Lambda$)
cannot be motivated on field theory grounds only.

It is interesting to mention that imposing on the field theory 
result  (\ref{pwl}) one of the string symmetries  $T \leftrightarrow
U$ or $T\leftrightarrow 1/T$, enables one to 
recover the {\it full} heterotic string result  \cite{Dixon:1990pc}  from 
that derived using only field theory methods. 
Thus one may obtain full string results by using only
 field theory methods supplemented  
by  some of the symmetries of the string, not respected
by the field theory approach, but imposed on the final field theory result.
 For more details on the exact link 
with the heterotic  string without Wilson lines see 
\cite{Ghilencea:2002ff,Ghilencea:2002ak}. This ends our 
discussion 
on the corrections in the PT regularisation scheme 
and their relation to string theory.

\vspace{0.5cm} 

Comparing the results for $\Omega_i$  in the  three regularisation
schemes eqs.(\ref{drl2}) to (\ref{ptc_result}), one 
notices  that the finite (regulator independent) 
part of $\Omega_i$ is the same in all cases
which is a consistency check of the calculation.
An important point to mention is that the result in the PT scheme has the
constraint that the compactification radii be large (in UV cutoff
units). The results in the DR and ZR  schemes
show that  the finite part of the  
one-loop correction has the value found without such 
restrictions.

Regarding the divergent part of the one-loop corrections, this is 
effectively dictated by the
regularisation choice one has to make, in agreement with 
the symmetries of the
model. Our discussion above shows that for  two compact
dimensions the PT regularisation is
indeed appropriate in calculations seeking the link with their 
string counterparts.
Further,  the $\zeta$-function regularisation leads  to an UV 
divergence which 
is milder (logarithmic) than in the PT   scheme with possible 
phenomenological implications.  This is important because 
models with ``power-like'' regime require 
in general a  significant amount of fine-tuning \cite{Ghilencea:1998st}.
It is  difficult to justify, without the knowledge of the 
full higher dimensional theory, in which case the $\zeta$-function  
regularisation is the right choice. The results of
 eqs.(\ref{drl2}) to  (\ref{pwl}) 
generalise, in the presence of Wilson lines,
early results \cite{dd10} for the radiative corrections 
from  two compact dimensions.

The one-loop corrections  obtained   in the DR, ZR or PT schemes
have strong  similarities with their one-dimensional
counterparts, eqs.(\ref{drresult}) to   (\ref{ptc_res}) with $T_2 U_2$ and 
$\rho_{2,\sigma} -U \rho_{1,\sigma}$ of eqs.(\ref{drl2}) to
(\ref{ptc_result}) replaced  in the one-dimensional case
by $R\mu$ and $\rho_\sigma$  respectively,  while 
$\ln(\vartheta_1/\eta)$ has as counterpart 
in the one-dimensional case  the term  $\ln[\sin\pi(\rho+i \chi)]$.  
A similar term  appears in compactification on $G_2$ manifolds 
\cite{Friedmann:2002ty} suggesting that  this latter 
correction is rather generic.

We end with a remark on possible phenomenological
implications. The result for $\Omega_i$ has a  divergence
which  depends - as expected - on the regularisation choice. 
Since  this is a
non-renormalisable theory, a natural question is whether one can 
make a prediction  without the knowledge of the fundamental, 
underlying theory which would otherwise  dictate the regularisation 
to use. If the gauge group G after orbifolding is a
grand unified group  which is further 
broken by Wilson lines  to a 
SM-like group, the coefficient of the (regularisation
dependent) divergent terms found in $\Omega_i^*$  is the same for all 
group factors into which  $G$  is broken ($G$-invariant). 
If so, such UV divergent terms of $\Omega_i^*$  
can then be absorbed  into the  redefinition of the initial 4D  
tree level coupling of the  group\footnote{
The method of ``absorbing'' the divergences in the initial 
tree level coupling also exists  in heterotic string models 
\cite{Nilles:1997vk} where gauge universal, gravitational effects  are 
included in the tree-level  coupling, in addition to the dilaton, 
with the remark that this is actually dictated by the symmetries 
of the (tree level coupling of) the string.}
 G. The newly defined coupling  can be 
regarded as the 4D ``MSSM-like'' unified coupling.  Further, the
remaining, {\it finite} part of $\Omega_i^*$ brings  a splitting  
term to  this coupling, due to Wilson lines vev
$\rho_\sigma$, 
but independent on the UV cutoff (regularisation). 
Finally, the ``MSSM-like''  massless states not included so far would 
bring the usual  logarithmic correction (UV scale dependent).  
This raises the possibility of allowing MSSM-like logarithmic 
unification even for ``large'' compact dimensions, and the aforementioned
splitting of couplings would ``mimic'' (at a scale of the order of
the compactification scale) what could be seen from 
a 4D point of view  as further running\footnote{in a 4D  renormalisable
theory.} up to  a high unification scale, such as that of 
the  MSSM  ($\approx 2\times 10^{16}$ GeV)  or higher.

\section{Conclusions}


The general structure of radiative corrections to gauge couplings
was investigated in generic 4D  models with one and two dimensional 
 compactifications in the presence of Wilson lines.  
The analysis was based on the
following observation.  Although one-loop  corrections are 
dependent on the exact  field content of the model, for the
compactifications considered one can still perform in a general
case,  the  one-loop integral
and the infinite sums over (non-zero) Kaluza-Klein levels associated with a 
 given state, component of a multiplet.  This leaves the much 
simpler analysis  of  determining which states have associated 
Kaluza-Klein towers, to a model-by-model analysis.

The evaluation  of the one-loop radiative corrections from 
compact dimensions summed up the individual  effects of  non-zero-level 
Kaluza-Klein modes. Although the models are non-renormalisable, 
the calculation  was kept general by considering the radiative effects
in three regularisation schemes: dimensional, zeta-function and 
proper-time cutoff regularisations for the UV divergences
 and the exact link among  these results was investigated.
The results in DR and $\zeta$-function regularisation schemes 
are very similar with the notable difference
that the (UV) pole structure of the DR scheme ($1/\epsilon$)  
is not present in the $\zeta$-function regularisation. This applies to 
both one and two extra dimensions cases. 
In the  ZR scheme  for $\Omega_i$ only a logarithmic
 divergence in the UV cutoff scale is present.  This is  important, since 
it provides an amount of sensitivity  of the radiative corrections 
to this scale smaller than that of other regularisations, which may 
be relevant for phenomenology. In the
DR and ZR schemes the finite part of the
results  is valid for either large or small compactification radii, for both 
one and two compact dimensions cases.

In proper-time regularisation the leading divergences of the radiative
corrections are for one and two compact dimensions  linear and
quadratic in scale, respectively. The finite (regulator independent) 
part is the same  as in the DR and zeta-function regularisation, which is a 
strong consistency check of the  calculation. 
The result in the
proper-time regularisation  is only valid for ``large'' compactification 
radii (in UV cutoff units), constraint not shared by
the results in the  DR and ZR schemes.
The effect of zero-modes (whose existence
is model dependent) can easily be 
added to the  results we obtained. In specific  cases they may even 
cancel the divergence from the entire KK tower of 
{\it non-zero} modes.  Finally, we also discussed 
the cases when for special values of the background/twists $\rho_\sigma$, 
one obtains one-loop {\it finite} 
results for the corrections due to positive or negative modes alone.

There remains the  question of which regularisation scheme
to use in (non-renormalisable) models with compact dimensions.
Explicit calculations and
comparison with the (heterotic) string  show that
proper-time cut-off regularisation is in exact quantitative 
agreement with the limit of large compactification radii of the 
string results. This 
applies to the case of two  compact dimensions 
which contribute to the  radiative corrections to the gauge couplings.   
Therefore  this regularisation is  an appropriate  choice  for
computing radiative corrections {\it for the purpose of establishing the 
link} with results from string theory. However, this regularisation
may be of limited use in field theory since is not gauge invariant.  
For the case of one compact dimension the
lack of string results  prevents one from  making a similar statement, 
and the choice of 
regularisation should follow  the usual guidelines 
such as its compatibility with  the symmetries of the model.

We addressed the possibility of 
making phenomenological predictions which are independent of the
UV divergence of the radiative corrections 
which, in the case of a grand unified
group G broken by Wilson lines vev's/twist $\rho_\sigma$, can be absorbed 
in the redefinition of the tree level coupling. This leaves a splitting
of the couplings at the compactification scale possibly compatible
with what can be regarded in a 4D
 (renormalisable) theory as further
``running'' up to a high, MSSM-like unification scale.

The paper provides all the  technical details necessary  in models
with one and two compact dimensions which examine  the one-loop 
corrections to the gauge couplings from Kaluza-Klein
thresholds  in the presence of  Wilson lines.
Although we discussed only the dependence of the corrections
on the UV cutoff/regulator, the paper provides the technical results for
investigating the change of the gauge couplings with respect to the
 (momentum) scale $q$ as well. Extensive mathematical  details 
of  regularisations of 
integrals and series present in one-loop corrections due to compact
dimensions were provided in  the Appendix.
Our results can be applied with minimal changes 
to many one- and  two-dimensional orbifolds with Wilson lines, by
making appropriate  re-definitions of the 
parameters of the models, such as the compactification radii ($R$), 
the twist of the initial fields  with respect to the compact
dimensions or the Wilson lines vev's ($\rho$).

\section{Acknowledgements}

The author  thanks Graham Ross for  discussions on this topic
and   suggestions on a previous version of this paper.
He also thanks Stefan F\"orste, Joel Giedt,  Fernando Quevedo and  
Martin Walter for discussions on related topics. This work was 
supported by a post-doctoral research fellowship from 
the Particle Physics and Astronomy Research Council  PPARC (UK).



\newpage
\section{Appendix}
\appendix
\def\theequation{\thesubsection-\arabic{equation}} 
\def\thesubsection{A} 
We provide general results for series of integrals present in 
one-loop corrections to the gauge couplings, evaluated 
in  DR, $\zeta$-function and proper-time cut-off regularisations,
for one and two compact dimensions (see also  Appendix A-4 in 
\cite{Ghilencea:2003kt}).
Notation used: A   ``primed'' sum $\sum_{m}' f(m)$ is a sum 
over   $m\in\bZ-\{0\}$;  $\sum_{m,n}' f(m,n)$ is a sum over all 
pairs of integers  $(m,n)$  {\bf excluding}  $(m,n)\!=\!(0,0)$.

\vspace{0.2cm}
\subsection{One compact dimension in Dimensional Regularisation  (DR)}
\setcounter{equation}{0}
\label{computing_1}

\noindent
$\bullet$ {\bf A.(1)}. 
\,\,\,\, We compute the following integral
\begin{equation}\label{drint}
\cR_\epsilon \equiv  \int_{0}^{\infty} \frac{dt}{t^{1+\epsilon}} 
\sum'_{m\in\bZ}
            e^{-\pi\,t\, [(m+\rho)^2 \beta+\delta]},
\qquad\quad \delta\geq 0,\,\,\beta\!>\! 0.
\end{equation}
$\Omega_i$ of eq.(\ref{drresult}) is then given by   
\begin{equation}\label{dfdr}
\Omega_i\Big\vert_{DR}= \beta_i(\sigma)/(4\pi)\,\,
\cR_\epsilon(\beta\!\ra\! 1/(R \mu)^2; \,\delta\!\ra\! (\chi/\mu)^2;\,
\rho\ra \rho_\sigma)
\end{equation}

\vspace{0.1cm}
\noindent
{\it Proof:} Consider first  $0<\delta/\!\beta\!\leq 1$. With the notation
$\rho= [\rho]+\Delta_\rho$,  $[\rho]\in\bZ$, $0\leq\!\Delta_\rho\!<1$
one has
\begin{eqnarray}
\!\!\!\!\!
\cR_\epsilon
& = &
\int_{0}^{\infty}\frac{dt}{t^{1+\epsilon}}
\bigg[
-e^{-\pi\,t\,\beta \rho^2}+e^{-\pi\,t\,\beta \Delta_\rho^2}
+\sum_{n\in\bZ}' e^{-\pi\, t\,(n+\Delta_\rho)^2\beta}\bigg]\,
e^{-\pi\, t\, \delta}
\nonumber\\
\nonumber\\
&=&\Gamma[-\epsilon] \pi^\epsilon
\bigg\{  (\delta+\beta \Delta_\rho^2)^\epsilon-
(\delta+\beta\rho^2)^\epsilon+
\bigg[\sum_{n>0}^{} [ \beta (n+\Delta_\rho)^2 
            +\delta]^\epsilon+(\Delta_\rho\ra -\Delta_\rho)\bigg]
            \bigg\}
\nonumber\\
\nonumber\\
&=&
\Gamma[-\epsilon] \pi^\epsilon
\big[ (\delta+\beta \Delta_\rho^2)^\epsilon-
(\delta+\beta\rho^2)^\epsilon]+\Gamma[-\epsilon] (\pi\beta)^\epsilon 
\bigg[
\zeta[-2\epsilon,1+\Delta_\rho]+\zeta[-2\epsilon,1-\Delta_\rho]\bigg]
\nonumber\\
\nonumber\\
&+& (\pi\beta)^\epsilon
\sum_{k\geq 1}\frac{\Gamma[k-\epsilon]}{k!}
\bigg[\frac{-\delta}{\beta}\bigg]^k\,\bigg\{
\zeta[2k-2\epsilon,1+\Delta_\rho]+(\Delta_\rho\ra-\Delta_\rho)\bigg\},
\quad 0<\delta/\beta\leq 1\label{eqa2}
\end{eqnarray}
which is convergent under conditions shown.
In the last step  we  used the {\it binomial} expansion~\cite{elizalde} 
\begin{equation}\label{zp}
\sum_{n\geq 0}[a(n+c)^2+q]^{-s}=a^{-s}\sum_{k\geq 0}
\frac{\Gamma[k+s]}{k\,!\, \Gamma[s]} \bigg[\frac{-q}{a}\bigg]^k
\zeta[2 k+ 2 s, c],\qquad 0<q/a\leq 1
\end{equation}
Here  $\zeta[q,a]$ with  $a\not=0,-1,-2,\cdots$ is the Hurwitz zeta
 function, (with $\zeta[q,a]=\sum_{n\geq 0} (a+n)^{-q}$ 
for $\textrm{Re}(q)>1$). 
Hurwitz zeta-function has one singularity (simple pole) at $q=1$
and  $\zeta[q,1]=\zeta[q]$ with $\zeta[q]$ the Riemann zeta function.
Further, using  eq.(\ref{eqa2}) and  the identity

$$\zeta[q,a]=a^{-q}+\zeta[q,a+1]$$ we obtain that
\begin{eqnarray}\label{eqa4}
\!\!\!\!\!\!\!\!\!\!\!\!\!\!\!\!
\cR_\epsilon\!\!&=& \!\pi^\epsilon \Gamma[-\epsilon] 
\big[ (\delta+\beta \Delta_\rho^2)^\epsilon-
(\delta+\beta\rho^2)^\epsilon] + (\pi\beta)^\epsilon \Gamma[-\epsilon]
\bigg[\zeta[-2\epsilon,1+\Delta_\rho]+\zeta[-2\epsilon,1-\Delta_\rho]\bigg]
\nonumber\\
\nonumber\\
&+&  \sum_{k\geq 1}
\bigg[\frac{-\delta}{\beta}\bigg]^k \frac{1}{k}
\bigg[\zeta[2k,1+\Delta_\rho]+\zeta[2k,2-\Delta_\rho] +
(1-\Delta_\rho)^{-2 k}\bigg]
\nonumber\\
\nonumber\\
&=&
\pi^\epsilon \Gamma[-\epsilon] 
\big[ (\delta+\beta \Delta_\rho^2)^\epsilon-
(\delta+\beta\rho^2)^\epsilon] + (\pi\beta)^\epsilon \Gamma[-\epsilon]
\bigg[\zeta[-2\epsilon,1+\Delta_\rho]+\zeta[-2\epsilon,1-\Delta_\rho]\bigg]
\nonumber\\
\nonumber\\
&-&\!\ln \frac{\vert \sin \pi[\Delta_\rho+i(\delta/\beta)^{\frac{1}{2}}]
\vert^2}{\pi^2 (\Delta_\rho^2+\delta/\beta)}
- 2
\ln\bigg[\Gamma[1-\Delta_\rho]\,\Gamma[1+\Delta_\rho]\bigg]
\end{eqnarray}
provided that
\begin{equation}\label{restrictions}
0<\frac{\delta}{\beta}\leq 1; \qquad  \frac{\delta}{\beta}< 
(1+\Delta_\rho)^2,
\qquad  \frac{\delta}{\beta} < (2-\Delta_\rho)^2
\end{equation}
Since $\Delta_\rho\!<\! 1$ we conclude that 
(\ref{eqa4}) is valid if the  first condition  (the
strongest) is respected:
\begin{equation}\label{rs2}
0<\frac{\delta}{\beta}\leq 1;
\end{equation} 
In the last step of deriving 
eq.(\ref{eqa4})  for each of the series in Zeta functions  we used 
\cite{srivastava}
\begin{equation}\label{wrtqy}
\sum_{k\geq 1}\frac{t^{2k}}{k}\,\zeta[2 k,a] =
\ln\frac{\Gamma[a+t]\,\Gamma[a-t]}{\Gamma[a]^2}; \qquad |t|<|a|
\end{equation}
with $t=i(\delta/\beta)^\frac{1}{2}$, $a=1+\Delta_\rho,\, 2-\Delta_\rho$ 
and from which the last two  conditions in (\ref{restrictions}) emerged.
Finally, in the last step in (\ref{eqa4}) we also used
\begin{equation}\label{id}
\prod_{\pm} 
\Gamma[1\pm x\pm i y]= \frac{\pi^2 (x^2+y^2)}{\vert \sin \pi(x+i
y)\vert^2 },\qquad x,y \,\,\,\textrm{real}.
\end{equation}
where the product runs  over all 4 combinations  of plus/minus
signs in the argument of $\Gamma$ functions. Eq.(\ref{id})
can be easily proved using that $\Gamma[1-z]\, \Gamma[1+z]= \pi
z/\sin \pi z$.

In eq.(\ref{eqa4}) we now evaluate the $\epsilon$ dependent part for
$\epsilon\ra 0$  by using (see for example \cite{gr})
\begin{eqnarray}\label{rwq}
\zeta[-2\epsilon,q]& =& \frac{1}{2} -q-2 \epsilon
\,\frac{d}{dz}\zeta[z,q]\vert_{z=0}+\cO(\epsilon^2)\nonumber\\
\frac{d}{dz}\zeta[z,q]\vert_{z=0}& = &  \ln\Gamma[q]-\frac{1}{2}
\ln(2\pi),\qquad 
\Gamma[-\epsilon]  = 
-\frac{1}{\epsilon}-\gamma+\cO(\epsilon),\nonumber\\
x^\epsilon &  = & 1+\epsilon\ln x+\cO(\epsilon^2)
\end{eqnarray}
We finally find from eq.(\ref{eqa4}), (\ref{rs2}), (\ref{rwq})  that 
(if $\rho\in \bZ^*$, $\delta=0$ is excluded)
\begin{equation}\label{eq1}
\cR_\epsilon = \int_{0}^{\infty} \frac{dt}{t^{1+\epsilon}} \sum'_{m\in\bZ}
            e^{-\pi\,t\, [(m+\rho)^2 \beta+\delta]}
 =  \frac{1}{\epsilon}-\ln\frac{\vert 2\, \sin
\pi(\Delta_\rho+i(\delta/\beta)^{\frac{1}{2}})\vert^2}
{\pi e^\gamma\, \beta \, (\rho^2+\delta/\beta)},\qquad 0\leq 
\delta/\beta \leq 1
\end{equation}


\noindent
$\bullet$ {\bf A.(2)}.\,\,\,
We now evaluate $\cR_\epsilon$ 
for the case $\delta/\beta>1$ (with notation 
$\rho\equiv  [\rho]+\Delta_\rho$,  $[\rho]\in\bZ$, $0\leq\!\Delta_\rho\!<1$)
\begin{eqnarray}\label{rr1}
\cR_\epsilon  =   \int_{0}^{\infty} \frac{dt}{t^{1+\epsilon}} \sum'_{m\in\bZ}
            e^{-\pi\,t\, [(m+\rho)^2 \beta+\delta]}
 & = &
\int_{0}^{\infty}\frac{dt}{t^{1+\epsilon}}
\bigg[
\sum_{n\in\bZ} e^{-\pi\, t\,(n+\Delta_\rho)^2\beta}
-e^{-\pi\,t\,\beta \rho^2}\bigg]\,
e^{-\pi\, t\, \delta}
\nonumber\\
\nonumber\\
& = & \Gamma[-\epsilon] \pi^\epsilon
\bigg[
\sum_{n\in\bZ}^{} [ \beta (n+\Delta_\rho)^2 
            +\delta]^\epsilon
-(\delta+\beta\rho^2)^\epsilon 
 \bigg]
\end{eqnarray}
We further use the well-known expansion given below (for
details see for example (4.13) in \cite{elizalde}) 
\begin{eqnarray}
\!\!\!\!\!\!\!
\sum_{n\in\bZ} [a (n+c)^2+q]^{-s}&=&
\sqrt\pi \bigg[\frac{q}{a}\bigg]^{\frac{1}{2}}
\frac{\Gamma[s-1/2]}{\Gamma[s]}\, q^{-s}\nonumber\\
\nonumber\\
&+&
 \frac{4\pi^s}{\Gamma[s]} 
\bigg[\frac{q}{a}\bigg]^{\frac{1}{4}}
q^{-{s}/{2}} a^{-{s}/{2}}
\sum_{n=1}^{\infty} n^{s-1/2}\cos(2\pi n\,c)\, 
K_{s-1/2}(2\pi n \sqrt{q/a})\label{exp}
\end{eqnarray}
with $a\!>\!0$, $c\!\not=\!0,-1,-2,\cdots$ and 
which is rapidly convergent for $q/a> 1$. $K_w$ is the modified Bessel
function of index $w$. The first term
proportional to $q/a$  gives the leading contribution; the
remaining ones  give ``instanton-like'' corrections. This result
is then used to evaluate (\ref{rr1}). 
Compare  (\ref{exp}) rapidly convergent 
for $q/a> 1$ with (\ref{zp}) valid for $q/a< 1$.
Alternatively, instead of (\ref{exp}) 
one can simply use a Poisson re-summation in (\ref{rr1})
and the definition of the modified Bessel functions to reach the same
result.
With $s=-\epsilon$ in (\ref{exp})  and with  
\begin{equation}\label{klrwpt}
\Gamma[-\epsilon]  = -\frac{1}{\epsilon}-\gamma+\cO(\epsilon),\qquad
K_{-\frac{1}{2}}( z) =  \sqrt{\frac{\pi}{2 z}} e^{- z}
\end{equation}
one finds from (\ref{rr1}) and (\ref{exp})
\begin{equation}
\cR_\epsilon  =   \int_{0}^{\infty} \frac{dt}{t^{1+\epsilon}}
 \sum'_{m\in\bZ}
            e^{-\pi\,t\, [(m+\rho)^2 \beta+\delta]}
 =  \frac{1}{\epsilon}-\ln\frac{\left\vert\, 2\, \sin
\pi(\Delta_\rho+i(\delta/\beta)^{\frac{1}{2}})\right\vert^2}
{\pi e^\gamma\,(\delta+\beta\rho^2)},
\quad \textrm{if}\,\,\,\delta/\beta > 1
\label{eq2}
\end{equation}
To see the complementarity of (\ref{eq1}) and  (\ref{eq2}) note that the
latter is not valid for $\delta=0$ since (\ref{exp}) is not valid in
that case.

In conclusion from eqs.(\ref{eq1}), (\ref{eq2}) we have
that 
\begin{eqnarray}\label{drlast}
\cR_\epsilon & \equiv &  \int_{0}^{\infty} 
\frac{dt}{t^{1+\epsilon}} \sum'_{m\in\bZ}
            e^{-\pi\,t\, [(m+\rho)^2 \beta+\delta]}
= \frac{1}{\epsilon}-\ln\frac{\left\vert 2\, \sin
\pi(\rho+i(\delta/\beta)^{\frac{1}{2}})\right\vert^2}
{\pi e^\gamma\, (\beta \rho^2+\delta)},\quad \delta\!\geq\! 0,\,
\beta\!>\!0.
\nonumber\\
\nonumber\\
\nonumber\\
\cR^T_\epsilon
  & \equiv  &  \int_{0}^{\infty} \frac{dt}{t^{1+\epsilon}} \sum_{m\in\bZ}
            e^{-\pi\,t\, [(m+\rho)^2 \beta+\delta]}
=  -\ln\left\vert\, 2\, \sin
            \pi(\rho+i(\delta/\beta)^{\frac{1}{2}})\right
\vert^2,\quad\delta\!\geq\! 0, \,\beta\!>\!0.
\end{eqnarray} 
In  eq.(\ref{drlast}) we used the properties of the sine function 
to replace  $\Delta_\rho$ by $\rho$. 
The pole $1/\epsilon$ cancels between zero-mode and non-zero 
modes' contributions.  Eq.(\ref{drlast})  was used in the text
eq.(\ref{drresult}).


\noindent
$\bullet$ {\bf A.(3)}.\,\,\,\,\,
  We compute the integral
\begin{equation}\label{drint2}
\cR_\epsilon^{+} \equiv  \int_{0}^{\infty} 
\frac{dt}{t^{1+\epsilon}} \sum_{m>0}
            e^{-\pi\,t\, [(m+\rho)^2
            \beta+\delta]},\qquad\quad \delta\geq 0,\,\,\beta\!>\!0.
\end{equation}
which sums positive modes only.  $\cR_\epsilon^-$ 
which sums negative modes only is then 
$\cR_\epsilon^-=\cR_\epsilon^+(\rho\ra -\rho)$.
$\Omega_i^\pm$ mentioned  in the text, eq.(\ref{smr2}) and 
corresponding to summing only  positive (negative) Kaluza-Klein modes
is then given by  
\begin{equation}\label{pmmm}
\Omega_i^\pm\bigg\vert_{DR} \equiv
\beta_i(\sigma)/({4\pi})\,\, \cR_{\epsilon}^\pm
\Big(\beta\!\ra\! 1/(R \mu)^2; \,\delta\!\ra\! (\chi/\mu)^2;\,
\rho\ra \rho_\sigma\Big)
\end{equation}

\vspace{1cm}
\noindent
The calculation proceeds almost identically to {\bf A.(1)}. The result is:
\begin{eqnarray}\label{ddrr}
\cR_\epsilon^+ =\frac{1}{2\epsilon}
+\frac{\rho}{\epsilon}
+\ln\Big\vert \Gamma[1+\rho+i (\delta/\beta)^{1/2}]\Big\vert^2
-\ln(2\pi)+\Big[\frac12+\rho\Big]\ln (\pi \beta e^\gamma)
\end{eqnarray}
which shows that a new divergence $\rho/\epsilon$ is present.
One can easily verify that
\begin{eqnarray}
\cR_\epsilon^+ +\cR_\epsilon^-=\cR_\epsilon
\end{eqnarray}
with $\cR_\epsilon$ given in (\ref{drlast}). 
This shows that the divergence $\rho/\epsilon$ of separate 
contributions from the   positive and negative modes respectively 
is {\it cancelled  in their sum} which equals $\cR_\epsilon$. 
While $\cR_\epsilon$ corresponds to states propagating in 
both directions in
the compact dimension in the ``background'' $\rho$, 
$\cR_\epsilon^\pm$ account for effects propagating in one direction
only.

\vspace{1cm}
\noindent
Similar properties exist for the full one-loop radiative corrections
 $\Omega_i^\pm$ given
below,  corresponding to positive and negative modes respectively.
The radiative correction in DR due to  positive (negative) modes only 
 is 
\begin{eqnarray}\label{prtq}
\Omega_i^\pm\bigg\vert_{DR} & \equiv & \frac{\beta_i(\sigma)}{4\pi}\,
\cR_{\epsilon}^\pm
\Big(\beta\!\ra\! 1/(R \mu)^2; \,\delta\!\ra\! (\chi/\mu)^2;\,
\rho\ra \rho_\sigma\Big)
\nonumber\\
\nonumber\\
&=&
\frac{\beta_i(\sigma)}{4\pi}\,
\left\{
\frac{1}{2\epsilon} \pm \frac{\rho_\sigma}{\epsilon}+
\ln\Big\vert\Gamma(1 \pm \rho_\sigma+i \chi R)\Big\vert^2
-\ln(2\pi)+\Big[\frac{1}{2} \pm \rho_{\sigma}\Big]\ln \frac{\pi
  e^\gamma}{(R\mu)^2}
\right\}
\end{eqnarray}
One has that  $\Omega_i^+ +\Omega_i^- =\Omega_i$ with $\Omega_i$ as
in (\ref{drresult}). The  ``linear''
divergence $\rho_\sigma/\epsilon$ cancels between positive and
negative modes' contributions.

\newpage
\def\theequation{\thesubsection-\arabic{equation}} 
\def\thesubsection{B} 
\subsection{One compact dimension in  $\zeta$-function regularisation
(ZR).}
\setcounter{equation}{0}
\label{computing_zeta}


$\bullet$  Here we define/evaluate $\Omega_i$ of
eq.(\ref{zeta_result}) in the ZR scheme.
The one-loop correction to the gauge couplings in Zeta-function 
regularisation is defined by (proportional to) the derivative of the Zeta
function associated with the Laplacian on the compact manifold
and evaluated in 0. To see this note that  $\zeta$-function of 
the Laplacian 
(eigenvalues $\lambda_m>0$) is defined as
\begin{equation} \label{zt1}
\zeta_\Delta [s]\equiv \sum_{m}' \frac{1}{\lambda_m^s}
=\frac{1}{\Gamma[s]}\sum_{m}'
 \int_{0}^\infty \frac{dt}{t^{1-s}} \, e^{-\lambda_m\, t} 
\end{equation}
where we used that
\begin{equation}\label{gamma}
 Q^{-s} =\frac{1}{\Gamma[s]} \int_{0}^\infty \frac{dt}{t^{1-s}} \,
 e^{-Q\, t}, \qquad Q>0
 \end{equation} 
From (\ref{zt1}) the {\it formal}
derivative of the zeta function $\zeta_\Delta^{'}[0]$ 
is   an infinite sum of individual logarithms of 
$\lambda_m$. With $\lambda_m$ expressed in some mass units $\mu$, 
($\lambda_m = M_m^2/\mu^2$)   one has the {\it formal} result
\begin{eqnarray}\label{prwqz}
\frac{d \zeta_\Delta [s]}{d s}\bigg\vert_{s=0} 
= -\sum_{m}' \ln\lambda_m = \sum_{m}'\ln(\mu/M_m)^2
\end{eqnarray}
and the link of $\Omega_i$ with  the one-loop corrections is obvious; 
$\mu$ acts as  effective field theory UV cutoff.

From eq.(\ref{zt1}) we have
\begin{eqnarray}\label{zt2}
 \cR_\zeta \equiv\frac{d \zeta_\Delta [s]}{d s}\bigg\vert_{s=0}
 =  
\frac{d}{d s} \, \bigg[
\frac{1}{\Gamma[s]} \sum'_{m}
\int_{0}^\infty \frac{dt}{t^{1-s}}\,  e^{-\lambda_m
\,t}\bigg]_{s= 0}
\end{eqnarray}
which relates $\zeta$-function regularisation of an operator to its 
value in the DR scheme.
One can also include the contribution of the zero mode $m=0$ (if
$\lambda_0\not=0$)  in the definition of $\zeta_\Delta[s]$. 
Accordingly   $\cR_\zeta$ changes and  is relabelled  $\cR_\zeta^T$.

With $\lambda_m=(m+\rho)^2 \beta+\delta$ as general eigenvalues of 
Laplacian for one-dimensional case
(see eq.(\ref{massformula}))  with boundary conditions given in
the text, and using  the results of eq.(\ref{drlast}) 
\begin{eqnarray}\label{drlast2}
\cR_\epsilon & \equiv &  \int_{0}^{\infty} 
\frac{dt}{t^{1+\epsilon}} \sum'_{m\in\bZ}
            e^{-\pi\,t\, [(m+\rho)^2 \beta+\delta]}
= \frac{1}{\epsilon}-\ln\frac{\left\vert 2\, \sin
\pi(\rho+i(\delta/\beta)^{\frac{1}{2}})\right\vert^2}
{\pi e^\gamma\,(\delta+\beta\rho^2)}\nonumber\\
\nonumber\\
\nonumber\\
\cR^T_\epsilon
  & \equiv  &  \int_{0}^{\infty} \frac{dt}{t^{1+\epsilon}} \sum_{m\in\bZ}
     e^{-\pi\,t\, [(m+\rho)^2 \beta+\delta]}
=  -\ln\left\vert\, 2\, \sin
   \pi(\rho+i(\delta/\beta)^{\frac{1}{2}})\right
\vert^2
\end{eqnarray}
we finally find
\begin{eqnarray}\label{zetafl}
\!\!\!\!\!\!\!\!
\cR_\zeta\!\!&\!\! \!\! = \!\!\!\!&\!\! -
\frac{d}{d \epsilon}\bigg\{ \frac{\pi^{-\epsilon}}{\Gamma[-\epsilon]}
\, R_\epsilon\,
\bigg\}_{\epsilon=0}\!\!\!
=\! -\ln\frac{\vert 2\, \sin
\pi(\rho+i(\delta/\beta)^{\frac{1}{2}})\vert^2}
{(\delta+\beta\rho^2)}
\nonumber\\
\nonumber\\
\!\!\!\!\!\!\!\!\!\!\!
\cR_\zeta^T\!\! & \!\!\!\!  =\!\!\!\! & \!\! 
- \frac{d}{d\epsilon} 
\bigg\{ \frac{\pi^{-\epsilon}}{\Gamma[-\epsilon]} 
\, R_\epsilon^T\bigg\}_{\epsilon=0}\!\!\!\!
=\! -\ln\left\vert\, 
2\, \sin \pi(\rho+i(\delta/\beta)^{\frac{1}{2}})\right\vert^2
\end{eqnarray}
Comparing the results of the last two sets of equations, one notices
that (up to a constant) the result in $\zeta$-function regularisation 
is equal to  that in DR from which the  pole contribution was subtracted.

Eqs.(\ref{zt1}), (\ref{zt2}) (\ref{zetafl}) allow us 
to evaluate $\Omega_i$ of eq.(\ref{zeta_result}). 
This is given by 
\begin{equation}\label{rtpgl}
\Omega_i\bigg\vert_{ZR}\equiv {\beta_i(\sigma)}/(4\pi)\,\, \cR_\zeta \left(
\delta\ra \chi^2/\mu^2, \,\, \beta\ra 1/(R\mu)^2,\,\, 
\rho\ra \rho_\sigma\right)
\end{equation}
According to eq.(\ref{prwqz}) $\mu$ should be regarded
as the effective field theory  UV cutoff.

\vspace{1cm}
\noindent
$\bullet$  Using the DR results eq.(\ref{drint2}) 
of summing over positive (negative) modes only
\begin{eqnarray}
\cR_\epsilon^{\pm} & \equiv &  
\int_{0}^{\infty} \frac{dt}{t^{1+\epsilon}} \sum_{m>0}
   e^{-\pi\,t\, [(m\pm \rho)^2 \beta+\delta]} 
\nonumber\\
\nonumber\\
& = & 
\frac{1}{2\epsilon}
\pm \frac{\rho}{\epsilon}
+\ln\Big\vert \Gamma[1 \pm \rho+i (\delta/\beta)^{1/2}]\Big\vert^2
-\ln(2\pi)+\Big[\frac12 \pm \rho\Big]\ln (\pi \beta e^\gamma)
\end{eqnarray}
one finds the associated $\zeta$-regularised result for
positive (negative) modes' contribution
\begin{eqnarray}
\cR_\zeta^\pm   \equiv  -
\frac{d}{d \epsilon}
\bigg \{ \frac{\pi^{-\epsilon}}{\Gamma[-\epsilon]}
\cR_\epsilon^\pm 
\bigg \}
=
\ln\Big\vert \Gamma[1 \pm \rho+i (\delta/\beta)^{1/2}]\Big\vert^2
-\ln(2\pi)+\Big[\frac12 \pm \rho\Big]\ln \beta 
\end{eqnarray}
The effect of positive (negative) modes on the 
gauge couplings in  $\zeta$-function regularisation is then
\begin{eqnarray}\label{zrzr}
\Omega_i^\pm \bigg\vert_{ZR} & \equiv &  \frac{\beta_i(\sigma)}{4\pi} \,
\cR_\zeta^\pm \, \bigg(\delta\ra \chi^2/\mu^2;\, \beta\ra 1/(R\mu)^2; \,
\rho\ra \rho_\sigma\bigg)\nonumber\\
\nonumber\\
& = & \frac{\beta_i(\sigma)}{4\pi}\left\{
\ln\Big\vert \Gamma[1 \pm \rho_\sigma+i \chi R ]\Big\vert^2
-\ln(2\pi)+\Big[\frac12 \pm \rho_\sigma \Big]\ln\frac{1}{(R\mu)^2} \right\}
\end{eqnarray}
This result  was used  in eq.(\ref{smr1}).

\newpage
\def\theequation{\thesubsection-\arabic{equation}} 
\def\thesubsection{C} 
\subsection{One compact dimension in proper-time regularisation (PT).}
\setcounter{equation}{0}
\label{computing_2}


$\bullet$  Here we provide technical details used to derive the
result of
eq.(\ref{ptc_res}).
In the proper-time cutoff regularisation, 
the generic structure of the one-loop  corrections is
\begin{eqnarray}\label{prtw}
\cR_\xi & \equiv & \int_{\xi}^{\infty} \frac{dt}{t} \,\sum'_{n\in\bZ} 
e^{-\pi t\,[(n+\rho)^2 \, \beta +  \delta]},\qquad \xi\ra 0,\,\,\,
(\xi>0);\,\, \delta\geq 0,\,\,\beta> 0.
\end{eqnarray}
$\Omega_i$ of  eq.(\ref{ptc_res}) is then given by
\begin{equation}\label{ptpt}
\Omega_i\bigg\vert_{PT}\equiv \beta_i(\sigma)/(4\pi) \, \,\cR_\xi 
(\beta\ra 1/(R\mu)^2; \,\rho\!\ra\!
\rho_\sigma;\, \delta\!\ra\! \chi^2/\mu^2; \, \xi\ra\! \xi)
\end{equation}

\vspace{1cm}
\noindent
To obtain $\cR_\xi$ we use  eq.(A-9) of Appendix 
A-1 of \cite{Ghilencea:2003kt}. One has
\begin{eqnarray}
\cR_\xi & = & 
\int_{\beta\xi}^{\infty} \frac{d t}{t} \sum_{n\in\bZ}' 
e^{-\pi t [(n+\rho)^2 +\delta/\beta]}
\nonumber\\
\nonumber\\
& = & \ln \bigg[(\rho^2+\delta/\beta) \pi e^\gamma\bigg]-
\ln\frac{ e^{-2/\sqrt{\xi\beta}}}{\xi\beta}
-\ln\left\vert\, 2\, \sin
\pi(\Delta_\rho+i(\delta/\beta)^{\frac{1}{2}})\right\vert^2
\end{eqnarray}
with $\Delta_\rho$ defined after eq.(\ref{dfdr}) and which is valid if:
\begin{eqnarray}
\label{conditions}
\xi\beta\ll \bigg\{ 1; \frac{1}{\pi \delta/\beta};
\frac{1}{\pi(\rho^2+\delta/\beta)}\bigg\}
\end{eqnarray}

\noindent
One concludes that
\begin{eqnarray}\label{ptclast}
\!\cR_\xi &\! \equiv\! & \sum'_{n\in\bZ} \int_{\xi}^{\infty} 
\frac{dt}{t} \,e^{-\pi t\,[(n+\rho)^2 \, \beta +  \delta]}
 =  -
\ln\frac{ e^{-2/\sqrt{\xi\beta}}}{\xi}
\!-\!\ln\frac{\left\vert\, 2\, \sin
\pi(\rho\!+\! i(\delta/\beta)^{\frac{1}{2}})\right\vert^2}
{[\pi e^\gamma(\delta+\beta  \rho^2 )]}
\nonumber\\
\nonumber\\
\nonumber\\
\cR_\xi^T & \!\equiv\! & \sum_{n\in\bZ} \int_{\xi}^{\infty} \frac{dt}{t} \,
e^{-\pi t\,[(n+\rho)^2 \, \beta +  \delta]}
= \frac{2}{\sqrt{\xi\beta}}
-\ln\left\vert\, 2\,
\sin\pi(\rho+i(\delta/\beta)^{\frac{1}{2}})
\right\vert^2
\end{eqnarray}
with condition (\ref{conditions}).  
In the above equations we replaced $\Delta_\rho$ by $\rho$.

Note that adding the zero-mode to $\cR_\xi$
 does not cancel the leading linear divergence
unlike the cases of DR or ZR schemes!
To understand the differences among the various regularisation schemes 
it is useful to compare the above result
of  the PT regularisation eq.(\ref{ptclast}), (\ref{conditions})
with  that of DR regularisation eq.(\ref{drlast}),  and that of 
$\zeta$-function regularisation eq.(\ref{zetafl}).

Eq.(\ref{ptclast}) was used in the text, eq.(\ref{ptc_res}).

\newpage
\def\theequation{\thesubsection-\arabic{equation}} 
\def\thesubsection{D} 
\subsection{Two compact dimensions in Dimensional Regularisation (DR).}
\label{appendixC}
\setcounter{equation}{0}


$\bullet$ 
For two compact dimensions we evaluate the integral:
\begin{equation}\label{ldef}
L_\epsilon\equiv \int_{0}^{\infty} \frac{dt}{t^{1+\epsilon}} 
\sum_{m_1,m_2\in\bZ}' 
e^{-\pi t\, \tau  \vert m_2+\rho_{2}- U (m_1+\rho_{1})\vert^2 },\qquad
\quad \tau>0; \quad U=U_1+i U_2
\end{equation}
$\Omega_i$ of eq.(\ref{drl2}) is then  given by
\begin{eqnarray}\label{eqrt}
\Omega_i\bigg\vert_{DR}=\beta_i(\sigma)/(4\pi) \,\, L_\epsilon \big(
\tau\ra 1/(T_2 U_2),\, \rho_{i}\ra \rho_{i,\sigma}\big)
\end{eqnarray}

\vspace{0.5cm}
\noindent
{\it Proof:} To compute $L_\epsilon$ we  use the Poisson 
re-summation eq.(\ref{p_resumation}), so the integrand  of 
$L_\epsilon$ becomes
\begin{eqnarray}\label{sums}
\sum_{m_1,m_2}'e^{-\pi t \, \tau 
\vert m_2+\rho_{2}- U (m_1+\rho_{1})\vert^2 }
&=&\sum_{m_2}'e^{-{\pi\, t \,\tau}  |m_2+\rho_2-U\rho_1|^2}
+\sum_{m_1}'\sum_{m_2\in\bZ} e^{-{\pi\, t\, \tau 
|m_2+\rho_2-U(m_1+\rho_1)|^2}}
\nonumber\\
\nonumber\\
&=&
  \sum_{m_2}' e^{- \pi\, t\, \tau  |m_2+\rho_2-U \rho_1|^2}+
  \frac{1}{\sqrt{t\,\tau}} \sum_{m_1}'
e^{-\pi t \, \tau U_2^2  \, (m_1+\rho_1)^2}
\nonumber\\
\nonumber\\
&+&\frac{1}{\sqrt{t\,\tau}}
\sum_{m_1}'\sum_{\tilde m_2}' e^{-\frac{\pi {\tilde m_2}^2}{t\,\tau} 
-\pi t\,\tau  U_2^2 \, (m_1+\rho_1)^2+2 \pi i  {\tilde m_2} 
(\rho_2-U_1(\rho_1+m_1))}
\end{eqnarray} 
A prime on the double sum in the lhs indicates  that
the mode $(m_1,m_2)\!\not=\!(0,0)$ is excluded.  If $\rho_1$ is
non-integer  the three series  in the rhs of (\ref{sums}) 
can be integrated separately over $(0,\infty$) to find
\begin{eqnarray}
L_\epsilon &=& L_1+L_2+L_3,\qquad \qquad \textrm{where}
\nonumber\\
\nonumber\\
\!\!L_1&\equiv &\int_{0}^{\infty}\frac{dt}{t^{1+\epsilon}}
\sum_{m_2}' e^{- \pi\, t\,\tau  \vert m_2+\rho_2-U\rho_1\vert^2}
=
\frac{1}{\epsilon}-
\ln\vert\, 2 \sin \pi(\rho_2-U \rho_1)\, \vert^2
+ \ln \Big[ \pi \,\tau  e^\gamma \vert \rho_2- U \rho_1\vert^2\Big]
\nonumber\\
\nonumber\\
\!\!L_2&\equiv & \frac{1}{\sqrt{\tau}}
\int_{0}^{\infty}\!\frac{dt}{t^{3/2+\epsilon}}
\sum_{m_1}'e^{-\pi t\,\tau \, U_2^2\,  (m_1+\rho_1)^2}
= 2\pi  U_2 \Big[\vert \rho_1\vert+ \frac{1}{6} -\Delta_{\rho_1}
+\Delta_{\rho_1}^2\Big],
\nonumber\\
\nonumber\\
\!\!L_3&\equiv & \frac{1}{\sqrt{\tau}}
\int_{0}^{\infty}\!\frac{dt}{t^{3/2+\epsilon}} 
\sum_{m_1}'\sum_{\tilde m_2}' e^{-\pi \tilde m_2^2\frac{1}{t\,\tau}
-\pi t\,\tau  U_2^2(m_1+\rho_1)^2+2 \pi i \tilde m_2 (\rho_2-U_1
(\rho_1+m_1))} 
\nonumber\\
\nonumber\\
&=&\ln \Big \vert 2 \sin \pi (\rho_2-U \rho_1)\Big\vert^2
-2\pi  U_2  \Big[\vert \rho_1\vert+\frac{1}{6} -
\Delta_{\rho_1}\Big]
-\ln\bigg\vert 
\frac{\vartheta_1(\Delta_{\rho_2}-U\Delta_{\rho_1}\vert U)}{\eta(U)}
\bigg\vert^2 
\label{cL123}
\end{eqnarray}
where  $\Delta_y$ denotes the  positive definite 
fractional part of $y$  defined as $y=[y]+\Delta_y$, 
$0<\Delta_y<1$, with  $[y]$ an integer number. 
$\vartheta_1(z\vert \tau)$ and $\eta(U)$ are special functions 
defined in the Appendix,  eqs.(\ref{R3_3}).

To evaluate $L_1$ we  used eq.(\ref{eq2}) with the following 
replacements for the arguments of this equation:
$\beta\ra \tau$, $\rho\ra  \rho_2-U_1 \rho_1$ and 
$\delta\ra \tau U_2^2 \rho_1^2$. 
To compute $L_2$  we used the results
of Appendix A of ref.\cite{Ghilencea:2003kt}, eq.(A-22) or 
more generally eqs.(A-43), (A-45). 
Regarding  $L_3$,  taking the limit $\epsilon\ra 0$ is allowed under 
the integral before performing the integral itself or the two sums. This is
justified by technical calculations (not shown) which prove
that $L_3$ is bound by an expression  which has no poles in 
$\epsilon\ra 0$. This is actually expected because the integrand is well 
defined for $t\ra 0$ or $t\ra\infty$ when $\epsilon=0$.
After setting $\epsilon=0$ the integral equals that evaluated in 
eqs.(A-28)  to (A-31) in  Appendix (A-3) of ref~\cite{Ghilencea:2003kt}.

Adding together $L_1, \, L_2,\,L_3$ we find for $U=U_1+i U_2$, $\tau>0$
\begin{eqnarray}\label{drl}
L_\epsilon& \equiv& 
\int_{0}^{\infty} \frac{dt}{t^{1+\epsilon}} \sum_{m_1,m_2\in\bZ}'
e^{-\pi t\, \tau 
\vert m_2+\rho_{2}- U (m_1+\rho_{1})\vert^2 }
\nonumber\\
\nonumber\\
&=& \frac{1}{\epsilon} +
\ln [ 
\pi \,\tau  e^\gamma ]
-\ln\bigg\vert 
\frac{\vartheta_1(\Delta_{\rho_2}-U\Delta_{\rho_1}\vert U)}{
[\rho_2- U \rho_1]\, 
\eta(U)}
\bigg\vert^2  + 2\pi U_2 \Delta_{\rho_1}^2,\qquad \tau>0
\end{eqnarray}
Further, one can make the  replacement
$\Delta_{\rho_i}\!\ra\! \rho_i$, due to the identity given in 
eq.(\ref{idt}).
Eq.(\ref{drl}) was used in the text, eq.(\ref{drl2}).

Using the properties of $\vartheta_1$  (Appendix~\ref{appendixG})
one also finds an  interesting limit of $L_\epsilon$ for
 $\rho_1\!=\rho_2\!=0$:
\begin{equation}
L_\epsilon (\rho_{1,2}=0)\, \, \equiv\, 
\int_{0}^{\infty} \frac{dt}{t^{1+\epsilon}} \sum_{m_1,m_2 \in\bZ}'
e^{-\pi t\, \tau  \vert m_2- U m_1\vert^2 }
\, =\, \frac{1}{\epsilon} + \ln\big[ \tau  e^\gamma /(4\pi)\big]-
\ln\vert\eta(U)\vert^4,\quad \tau>0.
\end{equation}
in agreement with eq.(B-12) of ref. \cite{Ghilencea:2002ff}.
Note that  the contribution of the $(0,0)$ mode - if added 
to  $L_\epsilon$~-
 would cancel the pole $1/\epsilon$ and $\ln\tau \vert
\rho_2-U \rho_1\vert$ term above.


\vspace{0.8cm}
\noindent
$\bullet$
One important observation is in place here. To find the scale
dependence of the divergence ($1/\epsilon$) of $L_\epsilon$
in the DR scheme one can  introduce a small/infrared (mass)$^2$ parameter
$\mu^2 \delta$ ($\delta$~dimensionless, $\delta>0$)
 in addition to the (mass)$^2$ of the Kaluza-Klein states in
the exponent in eqs.(\ref{twodim}),~(\ref{ldef}). 
This amounts to multiplying the integrand in eq.(\ref{twodim}) by
$e^{-\pi t\,\delta \,\mu^2}$ or that in (\ref{ldef}) by $e^{-\pi t\,
  \delta}$.   After a long  algebra one obtains the
following change for $L_1$, $L_2$, $L_3$:
\begin{eqnarray}\label{l1l2l3}
L_1&\ra & L_1'=L_1,\qquad\qquad\qquad\quad 
\textrm{if}\quad \delta\ra 0\nonumber\\
L_2&\ra & L_2'=L_2 +\frac{\pi \delta}{\epsilon} \frac{1}{\tau U_2},\qquad\,
\textrm{if}\quad  \delta\ra 0\nonumber\\
L_3 & \ra & L_3'= L_3,\qquad\qquad\quad \qquad \textrm{if}\quad \delta\ra 0
\end{eqnarray}
As a result
\begin{eqnarray}\label{lpp}
L'_\epsilon &\equiv &     
\int_{0}^{\infty} \frac{dt}{t^{1+\epsilon}} 
\sum_{m_{1,2}\in\bZ}' 
e^{-\pi t\, \tau  \vert m_2+\rho_{2}- U (m_1+\rho_{1})\vert^2
  -\pi\delta t } 
\nonumber\\
\nonumber\\
& = &    L_1'+L_2'+ L_3' \,\,\,\,=\,\,\,\,
 L_\epsilon+\frac{\pi \delta}{\epsilon} \frac{1}{\tau U_2},\quad
 \textrm{with}\quad \delta\ra 0; \quad \delta,\tau>0; \quad U=U_1+i U_2
\end{eqnarray}
with $L_\epsilon$ given in (\ref{drl}). 
Therefore a  divergence 
is emerging $\delta/ (\tau U_2\epsilon)$, 
induced by the change of $L_2$. With $\tau=1/(T_2 U_2)$ 
the divergence is proportional to  $T_2/\epsilon$, and is quadratic in
mass, given the definition of $T_2$. It is similar to that
of proper-time regularisation ($T_2/\xi$), see Appendix \ref{ptc_appendix}.  
Note that $L_2'$ which brings in this term
  is a  contribution from both compact dimensions, 
as Kaluza-Klein modes' effects from one dimension and Poisson re-summed
Kaluza-Klein zero-modes of the second compact dimension.
Also note a particular and useful limit of eq.(\ref{lpp}), that with
$\rho_1=\rho_2=0$.


\vspace{1.cm}
\noindent
$\bullet$
For future reference we also give the result of computing
the integral:
\begin{eqnarray}
L^*_\epsilon \equiv \int_{0}^{\infty}\!\! \!\frac{dt}{t^{1+\epsilon}}\!\!
\sum_{m_1,m_2\in\bZ}' e^{-\pi\, t\, \tau \,
\vert U m_1 - m_2\vert^2 -\pi\, t \,
  \delta}\qquad
\tau>0;\,\,\delta\geq 0,\,\,\, U\equiv U_1+ i \, U_2.
\\
\nonumber
\end{eqnarray}
{\it Proof:} Following the steps in eq.(\ref{sums}) one has $L^*_\epsilon = 
L^*_1+L_2^*+L_3^*$
with 
\begin{eqnarray}
L^*_1&\equiv&\int_0^\infty \!\frac{dt}{t^{1+\epsilon}} \, 
\sum_{m_2}' e^{-\pi
  \,t\,\tau\, m_2^2-\pi \, \delta\,t}
=\frac{1}{\epsilon}-\ln\bigg[
\frac{\vert 2\sinh\pi(\delta/\tau)^{1/2}\vert^2}{\pi \,
 e^\gamma\, \delta}\bigg]
\nonumber\\
\nonumber\\
L^*_2&\equiv&\frac{1}{\sqrt \tau} \int_0^\infty\! 
\frac{dt}{t^{3/2+\epsilon}}
\,\sum_{m_1}' e^{-\pi\,t\,\tau \,U_2^2 \,m_1^2 -\pi\,\delta\,t}
\nonumber\\
\nonumber\\
&=&\frac{\pi U_2}{3}+\frac{\pi\, \delta}{\epsilon\,\tau U_2}+
\frac{\pi \,\delta}{\tau\, U_2}\ln \bigg[4\pi\, e^{-\gamma}\tau
  U_2^2\bigg]+
 2\sqrt\pi\, U_2\sum_{k\geq
  1}\frac{\Gamma[k+1/2]}{(k+1)!}\bigg[\frac{-\delta}{\tau
    \,U_2^2}\bigg]^{k+1}\zeta[2k+1]
\nonumber\\
\nonumber\\
L^*_3 &\equiv& \frac{1}{\sqrt \tau} \int_0^\infty \!
\frac{dt}{t^{3/2+\epsilon}}\sum_{m_1}'\sum_{\tilde m_2}'
e^{-\pi\, {\tilde m_2}^2 /(t\,\tau)-\pi \, t\, \tau\, U_2^2 m_1^2-2 i
\pi \tilde m_2 m_1 U_1-\pi\,\delta\,t}
\nonumber\\
\nonumber\\
&=&-\ln \prod_{m_1\geq 1}
{\bigg\vert 1- e^{-2\pi (\delta/\tau+U_2^2 m_1^2)^{1/2}}\,
 e^{2 i \pi U_1 m_1}\bigg\vert^4}
\end{eqnarray}
For $L_1^*$ we used eq.(\ref{drlast}), for $L_2^*$ see 
eq.(B-11) to (B-15) in  Appendix B of \cite{Ghilencea:2002ak}. 
For $L_3^*$ one may set $\epsilon=0$ (no poles at $t\ra 0$ or $t\ra
\infty$) and use the integral representation of Bessel function
$K_{1/2}$ with $K_{1/2}(z)$ given in (\ref{klrwpt}).
Adding together the above contributions one has
\begin{equation}
L^*_\epsilon=
\frac{1}{\epsilon}
+\frac{\pi \delta}{\epsilon}\frac{1}{\tau U_2}
-\ln\bigg[4\pi e^{-\gamma}\frac{1}{\tau}\,\vert \eta(U)\vert^4 \bigg]
+\frac{\pi \delta}{\tau  U_2}\ln(4\pi e^{-\gamma}\,\tau \, U_2^2)
-2\ln
\frac{\sinh\pi (\delta/\tau)^{\frac{1}{2}}}{\pi 
(\delta/\tau)^{\frac{1}{2}}}
+\cW\bigg(\frac{\delta}{\tau}\bigg)
\end{equation}
with the constraint $0<\delta\, \vert U\vert^2/(U_2^2 \tau)\leq 1$,
$0< \delta/(\tau U_2^2)\leq 1$. Also $\cW(y\ra 0)\ra 0$ and
is defined as
\begin{equation}
\cW(y) \equiv
2\sqrt\pi  U_2 \sum_{k\geq  1}
\frac{\Gamma[k+1/2]}{(k+1)!}\bigg[\frac{-y}{U_2^2}\bigg]^{k+1}\!\!\!
\zeta[2k+1] -
\ln\! \!\prod_{m_1\geq 1}
\frac{
{\vert 1- e^{-2\pi (y+U_2^2 m_1^2)^{1/2}+2 i \pi U_1
    m_1}\vert^4}}{{
\vert 1- e^{2 i \pi U m_1}\vert^{4} }}
\end{equation}

\vspace{0.2cm}
\def\theequation{\thesubsection-\arabic{equation}} 
\def\thesubsection{E} 
\subsection{Two compact dimensions in $\zeta$-function
  regularisation (ZR)}
\setcounter{equation}{0}
\label{computing_zeta_two}


$\bullet$ 
Here we derive the result for $\Omega_i$ of
 eq.(\ref{zeta_reg2}) in the text.
The one-loop correction to gauge couplings in $\zeta$-function 
regularisation is proportional to the derivative of 
$\zeta$-function associated with the Laplacian on the compact manifold,
evaluated in 0 (See also the discussion in Appendix \ref{computing_zeta}).

$\zeta$-function of the Laplacian (eigenvalues $\lambda_{m,n}>0$)  
is defined as
\begin{equation} \label{qwqw}
\zeta_\Delta[s]\equiv \sum_{m,n\in\bZ}' \frac{1}{\lambda_{m,n}^s}
=\frac{1}{\Gamma[s]}\sum_{m,n\in\bZ}'
 \int_{0}^\infty \frac{dt}{t^{1-s}} \, e^{-\lambda_{m,n}\, t} 
\end{equation}
where we used eq.(\ref{gamma}) and the ``primed'' sum excluded the
$(0,0)$ mode.
Note that  as in the one extra-dimension case, 
one can express $\lambda_{m,n}$
in some mass units $\mu$,  $\lambda_{m,n}=M_{m,n}^2/\mu^2$ 
and  one has that, {\it formally }
\begin{eqnarray}\label{wzprtghkjl}
\frac{d \zeta_\Delta[s]}{d s}\bigg\vert_{s=0}=-\sum_{m,n}' 
\ln\lambda_{m,n}=
\sum_{m,n}' \ln(\mu/M_{m,n})^2
\end{eqnarray} 
and one can see the  link of this derivative with  the one-loop 
radiative corrections, given by a sum over individual logarithmic
corrections, with $\mu$ acting  as the UV cutoff of the model. 
Up to a beta function coefficient, eq.(\ref{wzprtghkjl}) is also 
in agreement  with the formal expression in eq.(\ref{frm}).

From eq.(\ref{qwqw}) one has
\begin{eqnarray}
L_\zeta \equiv
\frac{d \zeta_\Delta[s]}{d s}\bigg\vert_{s=0}  = 
 \frac{d}{d s} \, \bigg[
\frac{1}{\Gamma[s]} \sum'_{m,n\in\bZ}
\int_{0}^\infty \frac{dt}{t^{1-s}}\,  e^{-\lambda_m
\,t}\bigg]_{s= 0}
\end{eqnarray}
which relates $\zeta$-function regularisation of an operator to 
(the derivative of) its  DR result.

\noindent
With general eigenvalues of 
Laplacian for the two-dimensional case (see eq.(\ref{mass2})) 
\begin{equation}
\lambda_{m,n}=\tau \vert (m_2+\rho_2)-U(m_1+\rho_1)\vert^2,\qquad \tau>0
\end{equation}
and using  $L_\epsilon$ of
eq.(\ref{drl}), one has
\begin{eqnarray}\label{zzrr}
L_\zeta & = & 
- \frac{d}{d \epsilon}\left\{ \frac{\pi^{-\epsilon}}{\Gamma[-\epsilon]}
L_\epsilon \right\}=
\ln [ \tau ]
-\ln\bigg\vert 
\frac{\vartheta_1(\Delta_{\rho_2}-U \Delta_{\rho_1}\vert U)}{
 [{\rho_2}- U {\rho_1}]\, \eta(U)}
\bigg\vert^2  + 2\pi U_2 \Delta_{\rho_1}^2
\end{eqnarray}
One can further  replace
$\Delta_{\rho_i}\!\ra\! \rho_i$, due to the identity in eq.(\ref{idt}).
The result in  $\zeta$-function regularisation is equal to  that in DR 
from which the contribution of the pole was subtracted.

\vspace{1cm}
\noindent
$\bullet$ Eq.(\ref{zzrr}) was used to evaluate $\Omega_i$ 
 in the text eq.(\ref{zeta_reg2}) with
\begin{equation}\label{thrw}
\Omega_i\bigg\vert_{ZR}=\beta_i(\sigma)/(4\pi)\,\, L_\zeta \left(
\tau\!\ra 1/(T_2 U_2), \, \rho_i\ra \rho_{i,\sigma}\right)
\end{equation}

\vspace{1.cm}
\def\theequation{\thesubsection-\arabic{equation}} 
\def\thesubsection{F} 
\subsection{Two compact dimensions in proper-time 
  regularisation (PT).}
\label{ptc_appendix}
\setcounter{equation}{0}


$\bullet$ In the PT regularisation one evaluates  
(see the Appendix in ref.\cite{Ghilencea:2003kt})
\begin{eqnarray}\label{ptc}
L_\xi& \equiv& \int_{\xi}^{\infty} \frac{dt}{t} \sum_{m_{1,2}\in\bZ}'
e^{-\pi t\, \tau \vert m_2+\rho_{2}- U (m_1+\rho_{1})\vert^2 },\qquad
\xi\ra 0,\,\,\,\,\xi>0,\,\,\,\tau>0.
\end{eqnarray} 
with $U\equiv U_1+i U_2$.

\vspace{0.5cm}
\noindent
Therefore $\Omega_i$ of  eq.(\ref{ptc_result}) is
\begin{eqnarray}\label{lpwtr}
\Omega_i\bigg\vert_{PT}&=&\beta_i(\sigma)/(4\pi) \, L_\xi \left(
\tau\ra 1/(T_2 U_2)\right)
\end{eqnarray}

\vspace{0.6cm}
\noindent
Using the results of the Appendix in ref.\cite{Ghilencea:2003kt}
one has 
\begin{eqnarray}
L_\xi & = & \left\{\frac{1}{\xi \tau U_2}+\ln\xi \right\}
+\ln\Big[ \pi e^{\gamma} \tau \Big]
- \ln\bigg\vert 
\frac{\vartheta_1(\Delta_{\rho_2}-U\Delta_{\rho_1}\vert U)}{
\vert \rho_2-U \rho_1\vert\,\, \eta(U)}
\bigg\vert^2\!\!\! + 2\pi \, U_2 \Delta_{\rho_1}^2
\end{eqnarray}
with the condition
\begin{equation}
\frac{1}{\tau\,\xi}\gg \left\{ U_2^2,1/U_2^2\right\}
\end{equation}
The (divergent) expression in the curly braces is  
corresponding to $1/\epsilon$ in the  DR result, eq.(\ref{drl}).
Finally
\begin{eqnarray}\label{ptcreg}
L_\xi \left(
\tau\ra 1/(T_2 U_2)\right)
& = & \frac{T_2}{\xi}
+\ln \frac{ \pi e^{\gamma}}{(T_2/\xi)\, U_2}- \ln\bigg\vert 
\frac{\vartheta_1(\Delta_{\rho_2}-U\Delta_{\rho_1}\vert U)}{(\rho_2-U
  \rho_1)\, \eta(U)}
\bigg\vert^2\!\!\! + 2\pi \,\, U_2 \Delta_{\rho_1}^2\\
\nonumber
\end{eqnarray}
with 
\begin{equation}\label{vevs}
\max \left\{\frac{1}{R_1}, \frac{1}{R_2 \sin\theta}, 
\frac{\vert\rho_1\vert}{R_1},
\frac{\vert\rho_2 - U \rho_1\vert}{R_2 \sin\theta}\right\}
 \ll   \Lambda
\end{equation}
\newline
which was derived in eq.(52) of \cite{Ghilencea:2003kt}.
Here $T_2=\mu^2 R_1 R_2 \sin\theta$, $U_2=R_2/R_1 \exp(i\theta)$
and $\Lambda^2\equiv \mu^2/\xi$. One can make the  replacement
$\Delta_{\rho_i}\!\ra\! \rho_i$, due to the identity
 given in eq.(\ref{idt}).

\newpage
\def\theequation{\thesubsection-\arabic{equation}} 
\def\thesubsection{G} 
\subsection{Mathematical Appendix, Definitions and Conventions.}
\label{appendixG}
\setcounter{equation}{0}


\noindent
$\bullet$ In the text we used the special function $\eta$
\begin{eqnarray}\label{ddt}
\eta(\tau) & \equiv & e^{\pi i \tau/12} \prod_{n\geq 1} (1- e^{2 i
\pi\tau\, n}),
\nonumber\\
\nonumber\\
\eta(-1/\tau) & = & \sqrt{-i \, \tau}\,\eta(\tau),
\qquad
\eta(\tau+1)=e^{i\pi/12}\eta(\tau)
\end{eqnarray}
$\bullet$  We also used  the Jacobi function  $\vartheta_1$
\begin{eqnarray}\label{th}
\vartheta_1(z\vert\tau)&\equiv & 2 q^{1/8}\sin (\pi z) \prod_{n\geq 1} 
(1- q^n) (1-q^n e^{2 i \pi z}) (1- q^n e^{-2 i \pi z}), \qquad 
q\equiv e^{2 i \pi \tau}
\nonumber\\
 \nonumber\\
  &=& \frac{1}{i}\sum_{n\in\bZ} (-1)^n e^{i\pi\tau (n+1/2)^2}\,
  e^{(2n+1)i\pi z} 
\end{eqnarray}
which has the properties 
\begin{eqnarray}
\vartheta_1'(0|\tau) &=& 2 \pi \eta^3 (\tau), \quad 
\vartheta_1'(0|\tau)\equiv
\partial \vartheta_1(\nu|\tau)/\partial \nu|_{\nu=0}
\nonumber\\
\vartheta_1(\nu\,|\,\tau+1) & =& e^{i\pi/4} \, \vartheta_1(\nu|\tau),
\nonumber\\
\vartheta_1(\nu+1\,|\,\tau)& = & -\, \vartheta_1(\nu|\tau),
\nonumber\\
\vartheta_1(\nu+\tau\,|\,\tau)& = & -\,   e^{-i\pi \tau -2 i \pi \nu} \, 
\vartheta_1(\nu|\tau)
\nonumber\\
\vartheta_1(-\nu/\tau\, |-1/\tau)& =& 
e^{i\pi/4} \tau^{1/2} \exp(i\pi\nu^2/\tau)
\, \,\vartheta_1(\nu|\tau)
\label{R3_3}
\end{eqnarray}
Our conventions for $\vartheta_1$ are those of ref.\cite{Green}.
 $\vartheta_1(z\vert\tau)$ above 
is equal to $\vartheta_1(\pi z\vert\tau)$ of  \cite{gr}, eq.8.180(2).

\vspace{0.55cm}
\noindent
$\bullet$ Using these properties one can show that
\begin{eqnarray}\label{idt}
-\ln \Big\vert \vartheta_1(\Delta_{\rho_2} - U\Delta_{\rho_1}\vert U
)\Big\vert^2
+2\pi U_2 \Delta_{\rho_1}^2
=
-\ln \Big\vert \vartheta_1(\rho_2- U \rho_1| U) \Big\vert^2
+2\pi U_2 \rho_1^2
\end{eqnarray}
where $\Delta_{\rho_i}$ is the fractional part of $\rho_i$ defined as 
$\rho_i=[\rho_i]+\Delta_{\rho_i}, \quad 
[\rho]\in \bZ,\quad 0\leq \Delta_{\rho_{i}}<1$.

\vspace{0.55cm}
\noindent
$\bullet$ Throughout the Appendix we used the
Poisson re-summation formula:
\begin{equation}\label{p_resumation}
\sum_{n\in Z} e^{-\pi A (n+\sigma)^2}=\frac{1}{\sqrt A} \sum_{\tilde
n\in Z} e^{-\pi A^{-1} \tilde n^2+2 i \pi \tilde n \sigma}
\end{equation}

\end{document}